\documentclass[floatfix,
superscriptaddress,aps,showpacs,amsmath,amssymb,prl,
longbibliography,
twocolumn,a4paper,10pt]{revtex4-2}
\usepackage{lmodern}
\usepackage{bm} 
\usepackage{caption}
\usepackage{newtxmath}
\usepackage{siunitx}
\usepackage{amsbsy}
\usepackage{braket}
\usepackage{graphicx}
\usepackage{subcaption}
\usepackage{placeins}

\usepackage{ragged2e}  
\usepackage{microtype} 

\usepackage{caption}
\usepackage{booktabs}
\usepackage{dcolumn}
\usepackage{comment}
\usepackage{xcolor}
\usepackage{soul} 
\usepackage[colorlinks,
linkcolor=blue,
anchorcolor=blue,
citecolor=blue,
filecolor=blue,
menucolor=blue,
runcolor=blue,
urlcolor=blue,
frenchlinks=true]{hyperref}

\usepackage[noabbrev, nameinlink]{cleveref}
\begin{document}
\title{Information-Scrambling-Enhanced Quantum Sensing Beyond the Standard Quantum Limit}

\author{Yangyang Ge}
\thanks{These authors contributed equally.}
\affiliation{National Laboratory of Solid State Microstructures, School of Physics, Nanjing University, Nanjing 210093, China}
\affiliation{Shishan Laboratory, Nanjing University, Suzhou 215163, China}
\affiliation{Jiangsu Key Laboratory of Quantum Information Science and Technology, Nanjing University, Suzhou 215163, China}

\author{Haoyu Zhou}
\thanks{These authors contributed equally.}
\affiliation{National Laboratory of Solid State Microstructures, School of Physics, Nanjing University, Nanjing 210093, China}
\affiliation{Shishan Laboratory, Nanjing University, Suzhou 215163, China}
\affiliation{Jiangsu Key Laboratory of Quantum Information Science and Technology, Nanjing University, Suzhou 215163, China}

\author{Wen Zheng}
\email{zhengwen@nju.edu.cn}
\thanks{These authors contributed equally.}
\affiliation{National Laboratory of Solid State Microstructures, School of Physics, Nanjing University, Nanjing 210093, China}
\affiliation{Shishan Laboratory, Nanjing University, Suzhou 215163, China}
\affiliation{Jiangsu Key Laboratory of Quantum Information Science and Technology, Nanjing University, Suzhou 215163, China}

\author{Xiang-Min Yu}
\affiliation{National Laboratory of Solid State Microstructures, School of Physics, Nanjing University, Nanjing 210093, China}
\affiliation{Shishan Laboratory, Nanjing University, Suzhou 215163, China}
\affiliation{Jiangsu Key Laboratory of Quantum Information Science and Technology, Nanjing University, Suzhou 215163, China}
\affiliation{Synergetic Innovation Center of Quantum Information and Quantum Physics, University of Science and Technology of China, Hefei, Anhui 230026, China}
\affiliation{Hefei National Laboratory, Hefei 230088, China}

\author{Wei Fang}
\affiliation{Shishan Laboratory, Nanjing University, Suzhou 215163, China}
\affiliation{Jiangsu Key Laboratory of Quantum Information Science and Technology, Nanjing University, Suzhou 215163, China}

\author{Zhenchuan Zhang}
\affiliation{Shishan Laboratory, Nanjing University, Suzhou 215163, China}
\affiliation{Jiangsu Key Laboratory of Quantum Information Science and Technology, Nanjing University, Suzhou 215163, China}

\author{Wanli Huang}
\affiliation{Shishan Laboratory, Nanjing University, Suzhou 215163, China}
\affiliation{Jiangsu Key Laboratory of Quantum Information Science and Technology, Nanjing University, Suzhou 215163, China}

\author{Xiang Deng}
\affiliation{National Laboratory of Solid State Microstructures, School of Physics, Nanjing University, Nanjing 210093, China}
\affiliation{Shishan Laboratory, Nanjing University, Suzhou 215163, China}
\affiliation{Jiangsu Key Laboratory of Quantum Information Science and Technology, Nanjing University, Suzhou 215163, China}

\author{Haoyang Cai}
\affiliation{National Laboratory of Solid State Microstructures, School of Physics, Nanjing University, Nanjing 210093, China}
\affiliation{Shishan Laboratory, Nanjing University, Suzhou 215163, China}
\affiliation{Jiangsu Key Laboratory of Quantum Information Science and Technology, Nanjing University, Suzhou 215163, China}

\author{Xianke Li}
\affiliation{National Laboratory of Solid State Microstructures, School of Physics, Nanjing University, Nanjing 210093, China}
\affiliation{Shishan Laboratory, Nanjing University, Suzhou 215163, China}
\affiliation{Jiangsu Key Laboratory of Quantum Information Science and Technology, Nanjing University, Suzhou 215163, China}
\author{Kun Zhou}
\affiliation{National Laboratory of Solid State Microstructures, School of Physics, Nanjing University, Nanjing 210093, China}
\affiliation{Shishan Laboratory, Nanjing University, Suzhou 215163, China}
\affiliation{Jiangsu Key Laboratory of Quantum Information Science and Technology, Nanjing University, Suzhou 215163, China}
\affiliation{Synergetic Innovation Center of Quantum Information and Quantum Physics, University of Science and Technology of China, Hefei, Anhui 230026, China}
\affiliation{Hefei National Laboratory, Hefei 230088, China}

\author{Hanxin Che}
\affiliation{National Laboratory of Solid State Microstructures, School of Physics, Nanjing University, Nanjing 210093, China}
\affiliation{Shishan Laboratory, Nanjing University, Suzhou 215163, China}
\affiliation{Jiangsu Key Laboratory of Quantum Information Science and Technology, Nanjing University, Suzhou 215163, China}

\author{Tao Zhang}
\affiliation{National Laboratory of Solid State Microstructures, School of Physics, Nanjing University, Nanjing 210093, China}
\affiliation{Shishan Laboratory, Nanjing University, Suzhou 215163, China}
\affiliation{Jiangsu Key Laboratory of Quantum Information Science and Technology, Nanjing University, Suzhou 215163, China}

\author{Lichang Ji}
\affiliation{National Laboratory of Solid State Microstructures, School of Physics, Nanjing University, Nanjing 210093, China}
\affiliation{Shishan Laboratory, Nanjing University, Suzhou 215163, China}
\affiliation{Jiangsu Key Laboratory of Quantum Information Science and Technology, Nanjing University, Suzhou 215163, China}

\author{Yu Zhang}
\affiliation{National Laboratory of Solid State Microstructures, School of Physics, Nanjing University, Nanjing 210093, China}
\affiliation{Shishan Laboratory, Nanjing University, Suzhou 215163, China}
\affiliation{Jiangsu Key Laboratory of Quantum Information Science and Technology, Nanjing University, Suzhou 215163, China}

\author{Jie Zhao}
\affiliation{National Laboratory of Solid State Microstructures, School of Physics, Nanjing University, Nanjing 210093, China}
\affiliation{Shishan Laboratory, Nanjing University, Suzhou 215163, China}
\affiliation{Jiangsu Key Laboratory of Quantum Information Science and Technology, Nanjing University, Suzhou 215163, China}

\author{Shao-Xiong Li}
\affiliation{National Laboratory of Solid State Microstructures, School of Physics, Nanjing University, Nanjing 210093, China}
\affiliation{Shishan Laboratory, Nanjing University, Suzhou 215163, China}
\affiliation{Jiangsu Key Laboratory of Quantum Information Science and Technology, Nanjing University, Suzhou 215163, China}
\affiliation{Synergetic Innovation Center of Quantum Information and Quantum Physics, University of Science and Technology of China, Hefei, Anhui 230026, China}
\affiliation{Hefei National Laboratory, Hefei 230088, China}

\author{Xinsheng Tan}
\email{tanxs@nju.edu.cn}
\affiliation{National Laboratory of Solid State Microstructures, School of Physics, Nanjing University, Nanjing 210093, China}
\affiliation{Shishan Laboratory, Nanjing University, Suzhou 215163, China}
\affiliation{Jiangsu Key Laboratory of Quantum Information Science and Technology, Nanjing University, Suzhou 215163, China}
\affiliation{Synergetic Innovation Center of Quantum Information and Quantum Physics, University of Science and Technology of China, Hefei, Anhui 230026, China}
\affiliation{Hefei National Laboratory, Hefei 230088, China}
\affiliation{Jiangsu Physical Science Research Center, China}

\author{Yang Yu}
\email{yuyang@nju.edu.cn}
\affiliation{National Laboratory of Solid State Microstructures, School of Physics, Nanjing University, Nanjing 210093, China}
\affiliation{Shishan Laboratory, Nanjing University, Suzhou 215163, China}
\affiliation{Jiangsu Key Laboratory of Quantum Information Science and Technology, Nanjing University, Suzhou 215163, China}
\affiliation{Synergetic Innovation Center of Quantum Information and Quantum Physics, University of Science and Technology of China, Hefei, Anhui 230026, China}
\affiliation{Hefei National Laboratory, Hefei 230088, China}
\affiliation{Jiangsu Physical Science Research Center, China}

\date{\today}

\begin{abstract}
Quantum sensing promises measurement precision beyond classical limits, but its practical realization is often hindered by decoherence and the challenges of generating and stabilizing entanglement in large-scale systems. 
Here, we experimentally demonstrate a scalable, scrambling-enhanced quantum sensing protocol, referred to as butterfly metrology, implemented on a cross-shaped superconducting quantum processor. By harnessing quantum information scrambling, the protocol converts local interactions into delocalized metrologically useful correlations, enabling robust signal amplification through interference of the scrambled and polarized quantum states.
We validate the time-reversal ability via Loschmidt echo measurements and quantify the information scrambling through out-of-time-ordered correlators, establishing the essential quantum resources of our protocol.
Our measurements reveal that the sensing sensitivity surpasses the standard quantum limit (SQL) with increasing qubit number, reaching 3.78 in a 9-qubit configuration, compared to the SQL of 3.0.
The scheme further exhibits inherent robustness to coherent control errors and probed signal noise.
This work demonstrates a readily scalable path toward practical quantum sensing advantages with prevalent experimental platforms.
\end{abstract}
                             
\maketitle

\section{Introduction}
Quantum mechanics has enabled fundamental enhancements in measurement precision, advancing the field of quantum metrology~\cite{Montenegro2025,Giovannetti2004,Degen2017,Giovannetti2006}. This approach leverages distinct quantum resources~\cite{HORODECKI2013,Garcia2023}, including superposition, entanglement and squeezing, to surpass classical measurement bounds.
Entangled states such as spin-squeezed state~\cite{Kitagawa1993,Wineland1992,Gross2012} and GHZ state~\cite{Greenberger1990,Bollinger1996,Mooney2021} provide canonical examples of metrological resources capable of achieving super-linear scaling of the Fisher information~\cite{Rao1992,Braunstein1994,Zhang2023} with respect to the resource consumption.
These capabilities underpin major advances in precision measurement, including timekeeping~\cite{Derevianko2011,Takamoto2005,Rosenband2008}, gravitational-wave detection~\cite{Schnabel2010,Abbott2016,Tse2019}, and test of fundamental physics.
As quantum sensing platforms grow in size and complexity, they inevitably encounter challenges stemming from decoherence, environmental coupling, and imperfect control. 
These issues become particularly severe in quantum many-body systems, where intricate dynamics and strong interactions make precise state preparation and readout notoriously difficult.
Traditionally, such complexity has been viewed as a roadblock to deploying quantum sensors at scale.

One promising approach is the use of time-reversal protocols~\cite{Colombo2022,Wang2024,Colombo2023,Linnemann2016,Geier2024}, which reverse the system's evolution after a perturbation and thereby offer intrinsic robustness to readout noise. Beyond time reversal alone, recent advances suggest that quantum information scrambling~\cite{Swingle2018,bhattacharyya2022,Campisi2017,Xu2024,Hayden2007,Sekino2008,lashkari2013} can be harnessed to further enhance sensing performance. It describes information spreading, where initially localized quantum information can be dispersed throughout the exponentionally large many degrees of freedom of the entire system. It is intrinsically linked to quantum chaos~\cite{Xu2024,Luca2016,Srednicki1999}, which quantify how rapidly local perturbations propagate through entanglement and operator growth~\cite{Nahum2017,Roberts2015,Swingle2018}. This phenomenon plays a fundamental role in understanding complex quantum dynamics, serving as key mechanism behind the thermalization in isolated quantum systems~\cite{Rigol2008,Kaufman2016,Neill2016}. 
Its study provides crucial insights into frontier problems in modern physics, including the fast-scrambling dynamics conjectured in black holes~\cite{Sekino2008,Maldacena2016} and the breakdown of thermalization in many-body localized (MBL) systems~\cite{Abanin2019}.
Far from being an obstacle, quantum information scrambling can be transformed into a resource, dynamically generating the metrologically useful entangled states~\cite{Li2023}, thereby enhancing quantum sensing capabilities. 

\begin{figure*}[ht]
  \centering
  \includegraphics[width=0.95\textwidth]{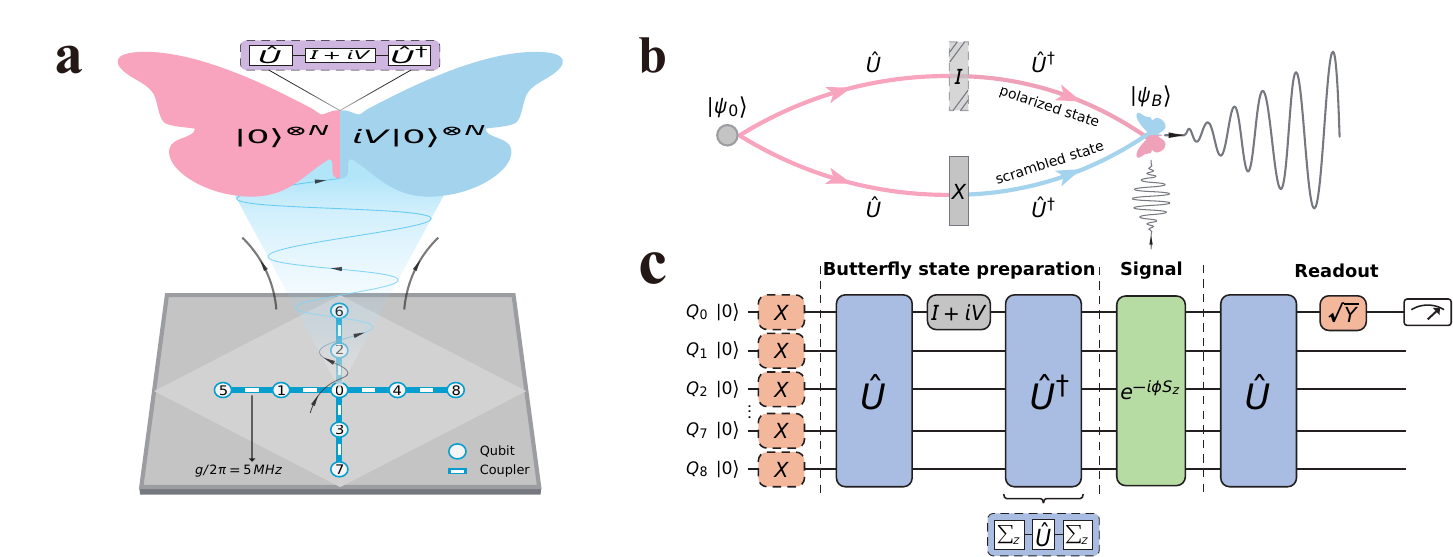}
    \caption{ \justifying 
    \textbf{Concept and protocol for quantum sensing enhanced by information scrambling.}
    \textbf{a,} Schematic of the experimental system and the butterfly quantum state. The lower panel shows the architecture of the 9-transmon superconducting quantum processor. The upper panel illustrates the butterfly state, an entanglement of a polarized state $\lvert 0 \rangle^{\otimes 9}$ and a scrambled state $\lvert \psi_{\textrm{sc}} \rangle$, engineered to approach a sensitivity scaling of half the Heisenberg limit.
    \textbf{b,} Interferometric representation of the protocol. The sensing signal $\phi$ is imprinted as a global phase on each branch of the butterfly state $\lvert \psi_{B} \rangle$, analogous to a quantum-enhanced Mach--Zehnder interferometer.
    \textbf{c,} The full metrological protocol: (i) Butterfly state preparation via a localized perturbation $(I + i V)/\sqrt{2}$ and forward unitary evolution $U(t)$, (ii) Signal acquisition during a sensing time $t$, and (iii) Readout via a rotation operation and projective measurement of the central qubit.
}
  \label{fig:1} 
\end{figure*}

In this work, we report the experimental demonstration of quantum-enhanced sensing via information scrambling, confirming the validity of the butterfly metrology protocol by Kobrin et al~\cite{Kobrin2024}.
By exploiting the exponential scrambling of a multiparticle system, this protocol generates metrologically useful entanglement between the scrambled state and the polarized state, enabling enhanced sensitivity beyond the SQL~\cite{Wineland1992,Giovannetti2004}. While the scrambling behavior of quantum chaotic systems has been widely explored~\cite{Landsman2019,Mi2021,Wang2021,Blok2021,Zhu2022,Li2023}, we adopt a systematic framework~\cite{Braumuller2022} to precisely define and quantify it. 
Using a 9-qubit superconducting processor~\cite{Kjaergaard2020,Krantz2019,Blais2021} arranged in a cross-shaped lattice, we present experimental observations of quantum information scrambing, measured via out-of-time-ordered correlator (OTOC)~\cite{Maldacena2016,Li2017,Braumuller2022,Abanin2025}. Second, we observe interference between the scrambled state and the polarized state, quantifying the scrambling-enhanced sensitivity for both local and global measurement protocols. Finally, we demonstrate the robustness of our protocols against coherent and incoherent noise resources.

\begin{figure*}[ht]
  \centering
  \includegraphics[width=0.95\textwidth]{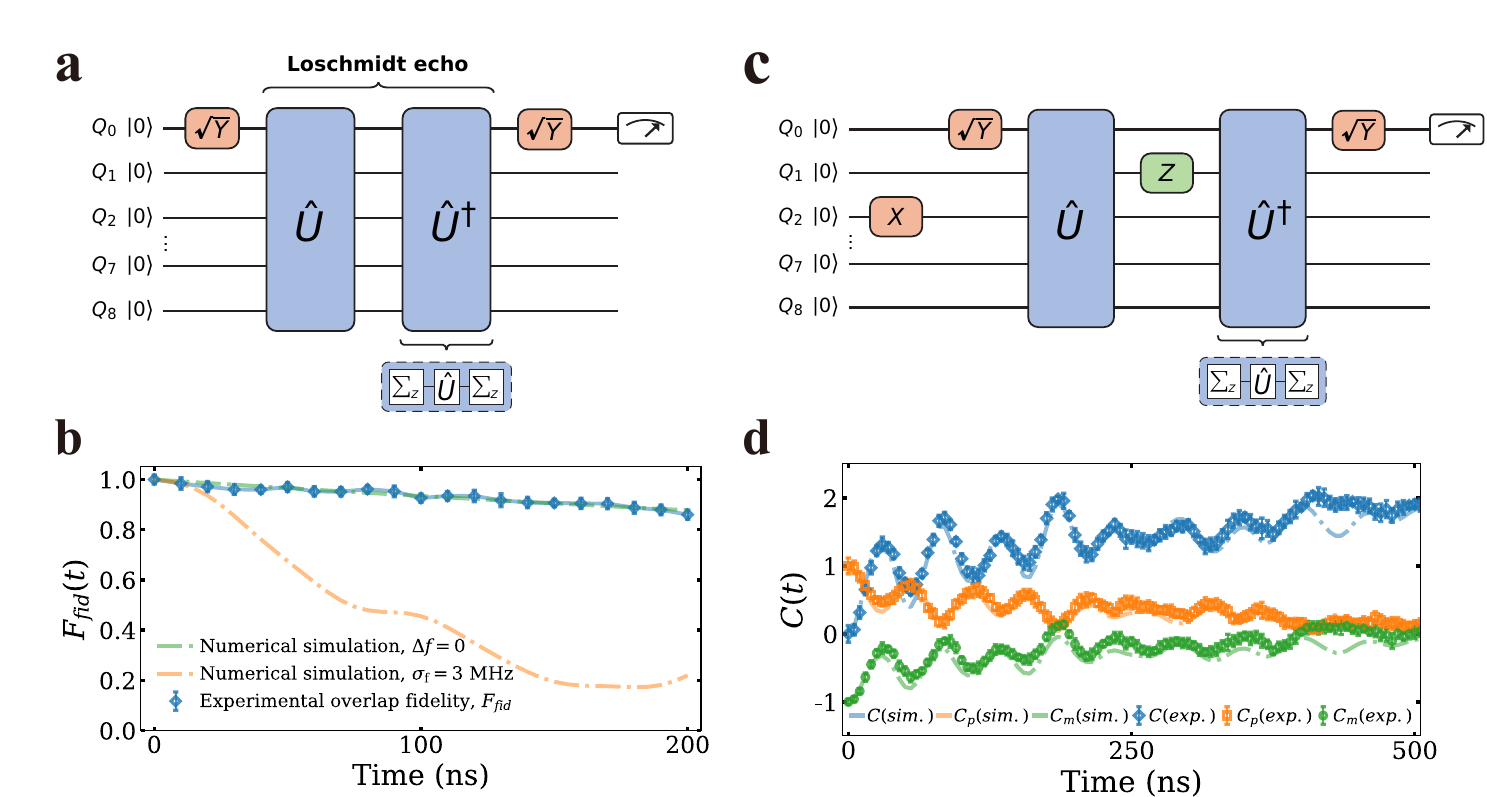}
\caption{ \justifying 
\textbf{a,} Quantum circuit for measuring the Loschmidt echo.
\textbf{b,} Experimental decay of the Loschmidt echo overlap fidelity, $F_{fid}(t)$, for $N=9$ qubits. Data points (diamonds) show experimental measurements with error bars. The solid line shows numerical simulation with zero frequency detuning ($\Delta f = 0~\mathrm{MHz}$), which shows excellent agreement with the experimental data. The dashed line shows simulation with Gaussian-distributed detunings ($\sigma_f = 3~\mathrm{MHz}$), demonstrating how frequency disorder perturbs the time-reversal evolution.
\textbf{c,} Quantum circuit for probing information scrambling via the OTOC.
\textbf{d,} Experimental measurement of the OTOC, $C(t)$, showing its characteristic decay at a evolution time of $500~\mathrm{ns}$, signaling the onset of scrambling. Data points show experimental measurements for different operator combinations $C$, $C_p$, and $C_m$, while dashed lines show corresponding numerical simulations.
}
  \label{fig:2} 
\end{figure*}

\section{Implementing the butterfly metrology scheme with superconducting qubits}
Traditional quantum-enhanced sensing approaches rely on carefully engineered entangled states like GHZ state or spin-squeezed state, which are challenging to prepare in many physical systems. The butterfly metrology protocol provides a universal framework for approaching half of the Heisenberg-limited sensitivity $(\eta \approx 2/N)$ by exploiting the natural scrambling dynamics of interacting quantum systems. 

The protocol begins by preparing a ``butterfly state", illustrated schematically in Fig.~\ref{fig:1}a. 
This state is an entanglement between two distinct branches: a fully polarized product branch \(\lvert 0 \rangle^{\otimes 9}\) and a strongly scrambled branch \(\lvert \psi_{\textrm{sc}} \rangle\).
As shown in the circuit diagram (Fig.~\ref{fig:1}c), it is constructed by applying a forward unitary evolution $U=e^{-i H t}$, introducing a localized perturbation $(I + i V)/\sqrt{2}$, and subsequently reversing the dynamics through $(U^\dagger)$. The resulting interference (Fig.~\ref{fig:1}b) between the ordered polarized branch and the disordered scrambled branch forms the basis for the metrological enhancement in our sensing protocol.
The sensitivity is improved by information scrambling and characterized by the OTOC,
 \begin{equation}
     \eta^{-1}_{\phi=0}=1/2 \sum_i (1-\bra{0}\sigma_i^z V(t)\sigma_i^z V(t)\ket{0}). 
 \end{equation} 
 As $V(t)$ spreads across the system, $V(t)$ and $\sigma_i^z$ become less commutated, operators grow and the OTOC decreases, enabling an improved sensitivity approaching $2/N$, which is the fully scrambled case.
The enhancement in sensitivity is governed by information scrambling, captured by the decay of the qubit-resolved OTOC, $C_i(t)=\bra{0}\sigma_i^z V(t)\sigma_i^z V(t)\ket{0}$.

\textit{\textcolor{blue}{Experimental setup}}---%
In the experiment we implement butterfly metrology on a cross-shaped superconducting quantum processor comprising nine transmon qubits with tunable nearest-neighbour couplers (see Fig.~\ref{fig:1}a). 
Our system implements a two-dimensional Bose-Hubbard model, described in the laboratory frame by 
\begin{equation}\label{BoseHubbard}
H = \sum_i \left[ \omega_i a^\dagger_i a_i + \frac{U}{2} a^\dagger_i a^\dagger_i a_i a_i \right] + \sum_{\langle i,j \rangle} J \left( a_i^\dagger a_j + a_i a_j^\dagger \right),
\end{equation}
where $a_{i,j}^\dagger(a_{i,j})$ is the creation(annihilation) operator, $\omega_i$ is on-site energy, $J$ is the nearest-neighbour hopping strength, and $U$ quantifies the anharmonicity of transmon qubits.

\begin{figure*}
    \centering
    \includegraphics[width=0.95\textwidth]{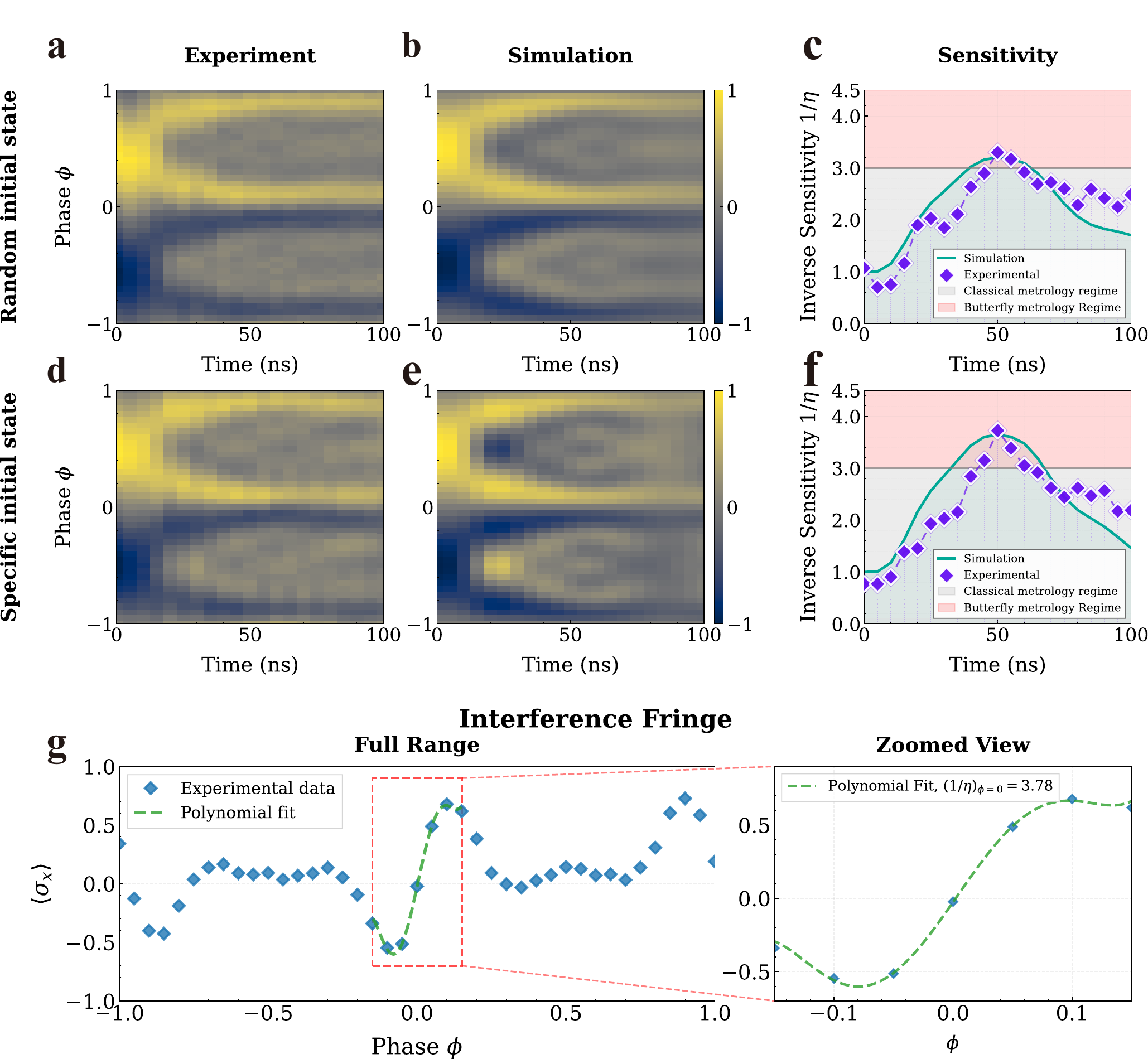}
     \caption{\justifying 
     \textbf{Quantum metrology in the 9-qubit configuration.}
\textbf{a,} \textbf{Random initial state: experimental interference pattern.} Measured local expectation value $\langle \sigma_x \rangle$ as a function of signal phase $\phi$ and evolution time $t$, quantifying interference between the scrambled quantum state and the polarized sensing state. Data correspond to an ensemble of random initial states. 
\textbf{b,} \textbf{Random initial state: numerical simulation.} Theoretically predicted $\langle \sigma_x \rangle$ from exact many-body simulation of the experimental protocol. 
\textbf{c,} \textbf{Random initial state: extracted sensitivity.} Inverse sensitivity $1/\eta$ as a function of time, derived from the interference pattern in panel \textbf{a}. The shaded regions indicate metrological performance relative to the standard quantum limit (SQL): below SQL (gray) corresponds to classical sensing, while above SQL (red) demonstrates quantum-enhanced sensing. 
\textbf{d--f,} \textbf{Specific initial state.} Same sequence as panels \textbf{a--c,} but for a deterministically prepared initial state, highlighting enhanced sensitivity. 
\textbf{g,} \textbf{Interference fringe.} Cross-section of pannel at the optimal time (50 ns), showing the amplified fringe curvature that yields maximal inverse sensitivity $1/\eta = 3.78$ (fourier fit, dashed line), surpassing the SQL.
}
     \label{fig:3}
\end{figure*}

\section{Quantum Loschmidt echo and Probing information propagation with OTOCs}

\textit{\textcolor{blue}{Time reversal and quantum Loschmidt echo}}---%
Fundamentally, the dynamics of a closed quantum system are unitary and thus inherently reversible, with the inverse of the evolution operator $U$ being its adjoint, $U^{\dagger}$. The cornerstone of our protocol is the practical application of this principle through the experimental realization of time-reversed dynamics.
This is accomplished by moving into a rotating frame and meticulously engineering an effective Hamiltonian that is the negation of the forward-evolution Hamiltonian. 

Starting from the Hamiltonian in the laboratory frame, we transform it into the rotating frame, obtaining the effective form 
\begin{equation}\label{RWF_Hamiltonian}
H = \sum_i \frac{\Delta\omega_i}{2} \sigma_i^z + \sum_{\langle i,j \rangle} J \left( a_i^\dagger a_j + a_i a_j^\dagger \right).
\end{equation}
In the Bose-Hubbard model, we find that $\sum_z a_i^\dagger a_j \sum_z = - a_i^\dagger a_j$ for any pair of adjacent qubits $\langle i,j \rangle$. Consequently, 
\begin{equation}
    \sum\nolimits_{z} H \sum\nolimits_{z} = \sum_{\langle i,j \rangle} -J_{ij} a_i^\dagger a_j = -H  
\end{equation}
The natural time evolution under this transformation satisfies
\begin{equation}
    \sum\nolimits_{z} U(t)\sum\nolimits_{z} = \sum\nolimits_{z} e^{-i H t} \sum\nolimits_{z} = e^{-i(-H)t} = U(-t),
\end{equation}
thus realizing the desired time-reversed evolution.

To benchmark the time reversibility of our lattice, we implement a quantum Loschmidt echo~\cite{Macri2016,Prosen2003,Karch2025} protocol. This involves letting the information from a sub-system disperse into the larger lattice, which acts as a bath, and then reversing the time evolution to observe the information's recovery. The circuit diagram for implementing the Loschmidt echo is depicted in Fig.~\ref{fig:2}a. Qubit $Q_0$ is initialized to the excited state $\ket{1}$, with the rest of the lattice in $\ket{0}$. The system evolves forward in time under $U(t)$, causing the local excitation to spread. The dynamics are then reversed by applying $U(-t)$. The success of the time reversal is evaluated by measuring the return probability of the excitation to qubit $Q_0$. 
The temporal decay of the Loschmidt echo is quantified by measuring the system's population in its initial state as a function of evolution time. As plotted in Fig.~\ref{fig:2}b, the echo fidelity decays from its ideal maximum of nearly 1.0, descending to a value of 0.8 within the $200$ ns of the time-reversed evolution. 
This marked drop captures the sensitivity of the quantum system to perturbations.

\textit{\textcolor{blue}{Probing quantum information scrambling with OTOCs}}---%
A standard method for studying information scrambling and characterizing the dynamics of interacting lattice systems in any dimension is to measure the OTOC. In our experiment, we probe these dynamics using the OTOC,
\begin{equation}
    F(t) = \left< W(t) V W(t) V \right>,
\end{equation}
where $V$ and $W$ are unitary local operators. To quantify operator growth with a Hermitian observable, we measure the expectation value of the squared commutator,
\begin{equation}
    C(t) = \left<[W(t),V]^\dagger [W(t), V]\right> = 2 - 2Re[F(t)].
\end{equation}
For our specific implementation, we choose $V=\sigma_j^z$ and $W=\sigma_i^x$, which gives 
\begin{equation}
    C = 2 - \bra{\psi} \sigma_j^z(t)\sigma_i^x\sigma_j^z(t)\sigma_i^x + \sigma_i^x\sigma_j^z(t)\sigma_i^x\sigma_j^z(t) \ket{\psi}. 
\end{equation}
The measured OTOC, depicted in Fig.~\ref{fig:2}d, reveals key dynamical features of the system. It exhibits persistent temporal oscillations, indicating that local information remains recoverable. At longer times, however, the OTOC displays a characteristic exponential decay. This decay quantifies the ``butterfly effect" and the rapid scrambling of quantum information. It is directly linked to the growth of operators and the Lyapunov exponent~\cite{Larkin1969,Maldacena2016,Scaffidi2019}, serving as a dynamical signature of quantum chaos. 

\begin{figure}
    \centering \includegraphics[width=8.5cm]{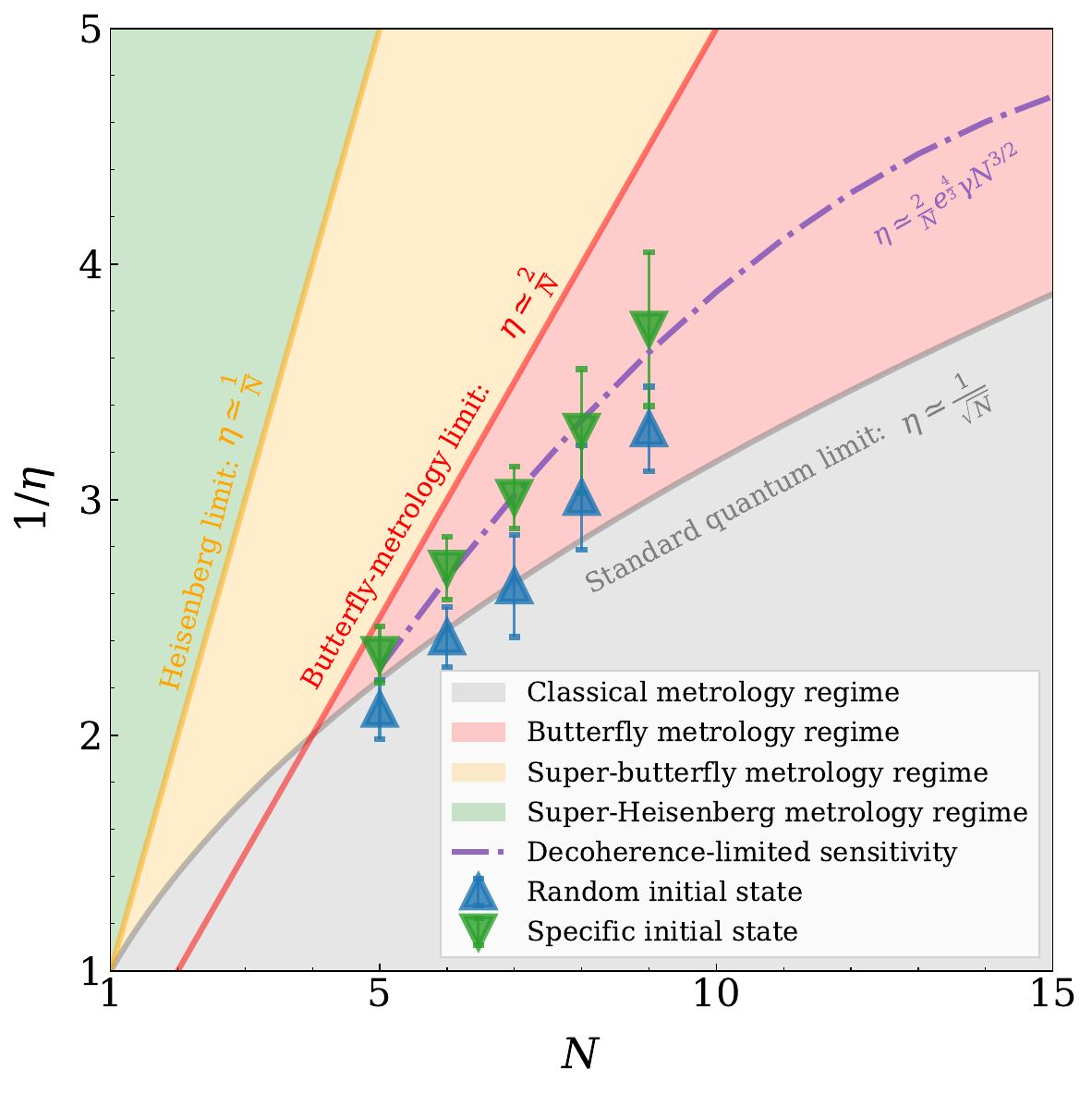} 
    \caption{\justifying 
    \textbf{Inverse sensitivity scaling of the butterfly metrology protocol.} The figure illustrates the scaling behavior of the inverse sensitivity as a function of the qubit number $N$, with the background divided into four distinct performance regimes: the classical metrology region (below the SQL), the butterfly metrology region (between the SQL and half of the Heisenberg limit), the super-butterfly metrology regime (between half of the Heisenberg limit and the Heisenberg limit), and the super-Heisenberg region (above the Heisenberg limit). The butterfly metrology limit at $\eta \simeq 2/N$ represents the theoretical upper bound for the butterfly metrology protocol. Experimental scaling is shown for two scenarios: the downward-oriented green triangle corresponds to measurements using a fixed initial state, while the upward-oriented blue triangle represents the average over randomly sampled initial states. Both cases exhibit favorable scaling beyond the SQL, approaching the butterfly metrology bound and demonstrating the metrological advantage enabled by scrambling dynamics.}
    \label{fig:4} 
\end{figure}

\section{Information scrambling enhanced sensitivity approaching half of the Heisenberg limit}
Our circuit for butterfly metrology is shown in Fig.~\ref{fig:1}c. It consists of three stages: butterfly state preparation, signal sensing, and readout. During the state preparation, a perturbation operator $iV$ disrupts the perfect cancellation between the forward evolution $U$ and its reverse $U^\dagger$. This generates a ``scrambled" state. This scrambled state and a simple polarized state $\ket{0}^{\otimes N}$ are then interfered. Finally, during signal sensing, these components acquire distinct relative phases in response to the signal. 
The measured interference pattern is shown in Fig.~\ref{fig:3}a for random initial states and Fig.~\ref{fig:3}d for a specific initial state. The corresponding numerical simulations are presented in Fig.~\ref{fig:3}b and e. To characterize the metrological sensitivity, we fit the curve of the local expectation value $\langle \sigma_x \rangle_\phi$ versus the signal strength $\phi$ and extract the inverse sensitivity $1/\eta$ as a function of evolution time $t$. As shown in Fig.~\ref{fig:3}c and f, the sensitivity surpasses the standard quantum limit of $\sqrt{N}$ (grey dashed line). In Fig.~\ref{fig:3}f, it further approaches half of the Heisenberg limit, scaling as $N/2$. For the specific initial state, the interference fringe at the optimal time (Fig.~\ref{fig:3}g) exhibits amplified curvature, yielding a maximal inverse sensitivity $1/\eta = 3.78$, which clearly exceeds the SQL.
\begin{figure*}
    \centering \includegraphics[width=0.95\textwidth]{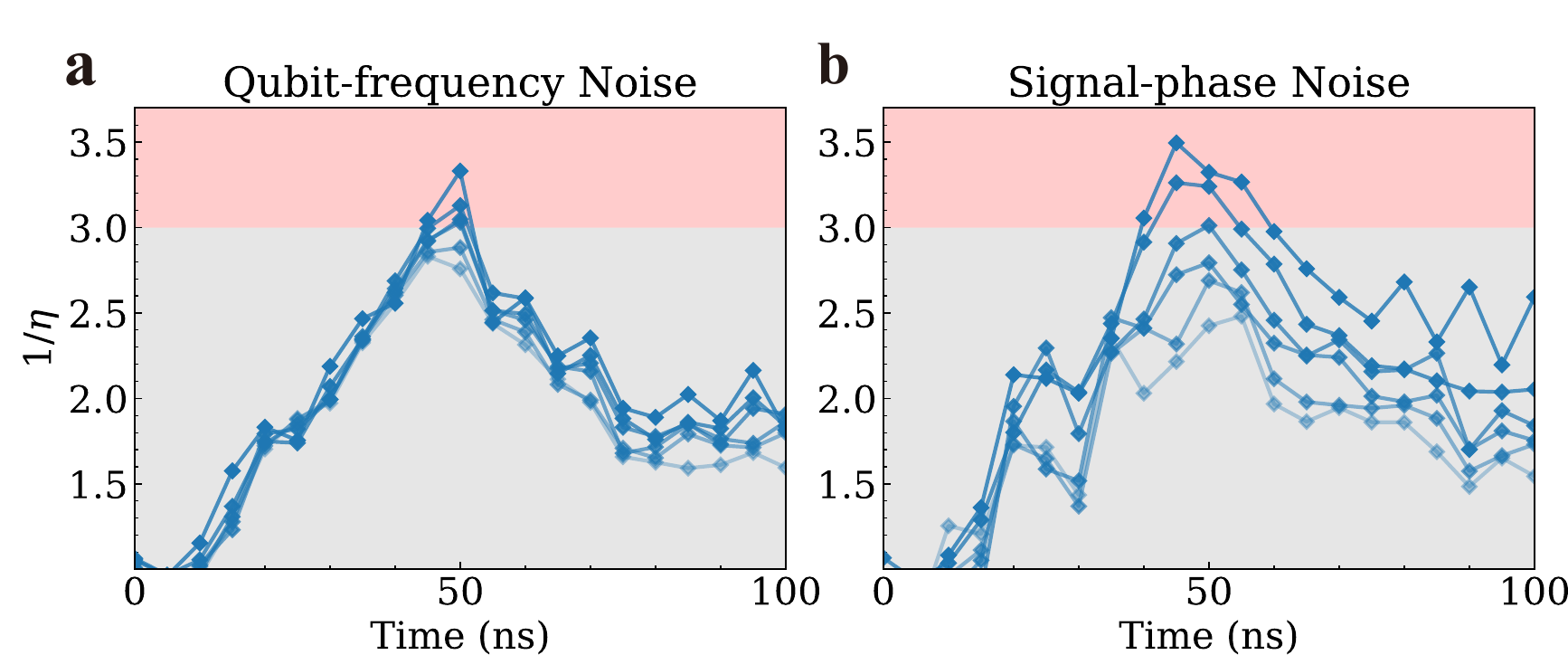} 
    \caption{\textbf{Measured inverse sensitivity under Gaussian noise.} The protocol's robustness is quantified by injecting noise during the time‑reversed state preparation and signal sensing stages. \textbf{a,} \textbf{Qubit‑frequency noise:} Inverse sensitivity $1/\eta$ versus $t$ for five values of $\sigma_\omega/2\pi$ (in MHz). \textbf{b,} \textbf{Signal‑phase noise:} Inverse sensitivity $1/\eta$ versus $t$ for five values of $\sigma_\phi$ (in rad). The protocol retains quantum enhancement (sensitivity above SQL) for noise strengths up to $\sigma_\omega/2\pi \approx 0.3\,\mathrm{MHz}$ and $\sigma_\phi \approx 0.2\,\mathrm{rad}$.}
    \label{fig:5}
\end{figure*}

In Fig.~\ref{fig:4}, we visualize the scaling behavior of the phase sensitivity as a function of system size, with different metrological regimes clearly delineated. The plot is structured to highlight the hierarchical landscape of quantum-enhanced sensing. The lower region represents performance bounded by classical strategies, where sensitivity scales no better than the SQL. Above this boundary lies the butterfly metrology regime, where the interplay between quantum chaos and controlled interference enables enhanced scaling -- yet still below the Heisenberg limit. The uppermost region is reserved for the super-Heisenberg regime, where sensitivities surpass even the Heisenberg limit, representing the ultimate quantum advantage in principle. Overlaying the background of the three distinct scaling regimes, we plot the experimentally extracted sensitivity values for butterfly metrology implemented on  superconducting quantum processors raging from 5 to 9 qubits. Two distinct trends are observed: one corresponding to experiments initialized in a fixed product state (shown in red) and another obtained from randomly sampled initial states (shown in blue). Both sets of data reside well within the butterfly metrology regime, reflecting the sub-Heisenberg scaling enabled by the scrambling-enhanced protocol. The sensitivity achieved with the fixed initial state consistently outperform that derived from random states, as expected from its optional constructive interference properties. 
However, neither trajectory reaches half of the Heisenberg limit, which represents the theoretical upper bound. This shortfall is primarily due to decoherence and other experimental imperfections that prevent full utilization of the quantum correlations intrinsic to the protocol. These results therefore chart the practical performance landscape of butterfly metrology under realistic noise conditions and establish a baseline for scaling toward larger quantum processors.
 
\textit{\textcolor{blue}{Robust to coherent and incoherent errors}}—
Our protocol demonstrates robustness against both coherent control errors and incoherent noise processes. To assess robustness against coherent errors, we introduce Gaussian-distributed fluctuations in the Hamiltonian parameters during the time-reversed state preparation and signal sensing stages, specifically by adding random variations to the qubit frequencies. To probe robustness against incoherent errors, we instead inject Gaussian noise directly into the sensed signal phase.

As shown in Fig.~\ref{fig:5}, the inverse sensitivity $1/\eta$ degrades only gradually with increasing noise strength in both cases, indicating that the protocol tolerates a substantial level of imperfections. In the absence of noise, the sensitivity clearly surpasses the SQL, demonstrating quantum enhancement. As the noise level increases, the enhancement is progressively reduced, and the sensitivity eventually falls below the SQL. Notably, quantum-enhanced performance persists up to noise strengths of approximately $\sigma_\omega/2\pi \approx 0.3~\mathrm{MHz}$ for qubit-frequency noise and $\sigma_\phi \approx 0.2~\mathrm{rad}$ for signal-phase noise, highlighting the intrinsic robustness of the protocol.

The detailed Gaussian noise model, including the statistical assumptions and implementation procedure used in the numerical simulations, is described in Supplementary Sec.~\textit{Gaussian Noise Model for Robustness Testing}.

\section{Conclusion}
In summary, we have experimentally realized the Loschmidt echo on a 9-qubit quantum chip, providing a direct verification of quantum information scrambling in a spin-chain-star model. Leveraging this complex forward and backward dynamics, we demonstrated a novel ``butterfly metrology" protocol that actively harnesses quantum information scrambling to enhance sensing capabilities. The implementation of this local protocol enabled a measurement sensitivity that not only surpasses the SQL but also approaches half of the Heisenberg limit. Furthermore, we have shown that our protocol exhibits inherent robustness against both coherent and incoherent errors, offering a significant practical advantage for mitigating ubiquitous experimental challenges such as low-frequency noise and calibration imperfections.
This work establishes a compelling connection between quantum information scrambling and practical metrological advantage and demonstrates the emerging scrambling-enhanced quantum sensing paradigm, offering a pathway to scalable, noise-resilient metrology in highly complex quantum systems.

Acknowledgments — We extend our appreciation to Bryce Kobrin for his insightful contributions during the discussions. Additionally, we thank Jiayu Ding and Orkesh Nurbolat for technical supports. This work was partially supported by the Quantum Science and Technology-National Science and Technology Major Project (Grant Nos.2021ZD0301702, 2024ZD0302000), Natural Science Foundation of Jiangsu Province (Grant Nos. BK20232002, BK20251248, BK20233001), National Natural Science Foundation of China (Grant Nos.U21A20436, 12074179, 12504579), and Natural Science Foundation of Shandong Province (Grant No. ZR2023LZH002).

\bibliography{main} 

\end{document}


\title{Supplementary Material for 
            ``Information-Scrambling-Enhanced Quantum Sensing Beyond the Standard Quantum Limit"}

\author{Yangyang Ge}
\thanks{These authors contributed equally.}
\affiliation{National Laboratory of Solid State Microstructures, School of Physics, Nanjing University, Nanjing 210093, China}
\affiliation{Shishan Laboratory, Nanjing University, Suzhou 215163, China}
\affiliation{Jiangsu Key Laboratory of Quantum Information Science and Technology, Nanjing University, Suzhou 215163, China}

\author{Haoyu Zhou}
\thanks{These authors contributed equally.}
\affiliation{National Laboratory of Solid State Microstructures, School of Physics, Nanjing University, Nanjing 210093, China}
\affiliation{Shishan Laboratory, Nanjing University, Suzhou 215163, China}
\affiliation{Jiangsu Key Laboratory of Quantum Information Science and Technology, Nanjing University, Suzhou 215163, China}

\author{Wen Zheng}
\email{zhengwen@nju.edu.cn}
\thanks{These authors contributed equally.}
\affiliation{National Laboratory of Solid State Microstructures, School of Physics, Nanjing University, Nanjing 210093, China}
\affiliation{Shishan Laboratory, Nanjing University, Suzhou 215163, China}
\affiliation{Jiangsu Key Laboratory of Quantum Information Science and Technology, Nanjing University, Suzhou 215163, China}

\author{Xiang-Min Yu}
\affiliation{National Laboratory of Solid State Microstructures, School of Physics, Nanjing University, Nanjing 210093, China}
\affiliation{Shishan Laboratory, Nanjing University, Suzhou 215163, China}
\affiliation{Jiangsu Key Laboratory of Quantum Information Science and Technology, Nanjing University, Suzhou 215163, China}
\affiliation{Synergetic Innovation Center of Quantum Information and Quantum Physics, University of Science and Technology of China, Hefei, Anhui 230026, China}
\affiliation{Hefei National Laboratory, Hefei 230088, China}

\author{Wei Fang}
\affiliation{Shishan Laboratory, Nanjing University, Suzhou 215163, China}
\affiliation{Jiangsu Key Laboratory of Quantum Information Science and Technology, Nanjing University, Suzhou 215163, China}

\author{Zhenchuan Zhang}
\affiliation{Shishan Laboratory, Nanjing University, Suzhou 215163, China}
\affiliation{Jiangsu Key Laboratory of Quantum Information Science and Technology, Nanjing University, Suzhou 215163, China}

\author{Wanli Huang}
\affiliation{Shishan Laboratory, Nanjing University, Suzhou 215163, China}
\affiliation{Jiangsu Key Laboratory of Quantum Information Science and Technology, Nanjing University, Suzhou 215163, China}

\author{Xiang Deng}
\affiliation{National Laboratory of Solid State Microstructures, School of Physics, Nanjing University, Nanjing 210093, China}
\affiliation{Shishan Laboratory, Nanjing University, Suzhou 215163, China}
\affiliation{Jiangsu Key Laboratory of Quantum Information Science and Technology, Nanjing University, Suzhou 215163, China}

\author{Haoyang Cai}
\affiliation{National Laboratory of Solid State Microstructures, School of Physics, Nanjing University, Nanjing 210093, China}
\affiliation{Shishan Laboratory, Nanjing University, Suzhou 215163, China}
\affiliation{Jiangsu Key Laboratory of Quantum Information Science and Technology, Nanjing University, Suzhou 215163, China}

\author{Xianke Li}
\affiliation{National Laboratory of Solid State Microstructures, School of Physics, Nanjing University, Nanjing 210093, China}
\affiliation{Shishan Laboratory, Nanjing University, Suzhou 215163, China}
\affiliation{Jiangsu Key Laboratory of Quantum Information Science and Technology, Nanjing University, Suzhou 215163, China}
\author{Kun Zhou}
\affiliation{National Laboratory of Solid State Microstructures, School of Physics, Nanjing University, Nanjing 210093, China}
\affiliation{Shishan Laboratory, Nanjing University, Suzhou 215163, China}
\affiliation{Jiangsu Key Laboratory of Quantum Information Science and Technology, Nanjing University, Suzhou 215163, China}
\affiliation{Synergetic Innovation Center of Quantum Information and Quantum Physics, University of Science and Technology of China, Hefei, Anhui 230026, China}
\affiliation{Hefei National Laboratory, Hefei 230088, China}

\author{Hanxin Che}
\affiliation{National Laboratory of Solid State Microstructures, School of Physics, Nanjing University, Nanjing 210093, China}
\affiliation{Shishan Laboratory, Nanjing University, Suzhou 215163, China}
\affiliation{Jiangsu Key Laboratory of Quantum Information Science and Technology, Nanjing University, Suzhou 215163, China}

\author{Tao Zhang}
\affiliation{National Laboratory of Solid State Microstructures, School of Physics, Nanjing University, Nanjing 210093, China}
\affiliation{Shishan Laboratory, Nanjing University, Suzhou 215163, China}
\affiliation{Jiangsu Key Laboratory of Quantum Information Science and Technology, Nanjing University, Suzhou 215163, China}

\author{Lichang Ji}
\affiliation{National Laboratory of Solid State Microstructures, School of Physics, Nanjing University, Nanjing 210093, China}
\affiliation{Shishan Laboratory, Nanjing University, Suzhou 215163, China}
\affiliation{Jiangsu Key Laboratory of Quantum Information Science and Technology, Nanjing University, Suzhou 215163, China}

\author{Yu Zhang}
\affiliation{National Laboratory of Solid State Microstructures, School of Physics, Nanjing University, Nanjing 210093, China}
\affiliation{Shishan Laboratory, Nanjing University, Suzhou 215163, China}
\affiliation{Jiangsu Key Laboratory of Quantum Information Science and Technology, Nanjing University, Suzhou 215163, China}

\author{Jie Zhao}
\affiliation{National Laboratory of Solid State Microstructures, School of Physics, Nanjing University, Nanjing 210093, China}
\affiliation{Shishan Laboratory, Nanjing University, Suzhou 215163, China}
\affiliation{Jiangsu Key Laboratory of Quantum Information Science and Technology, Nanjing University, Suzhou 215163, China}

\author{Shao-Xiong Li}
\affiliation{National Laboratory of Solid State Microstructures, School of Physics, Nanjing University, Nanjing 210093, China}
\affiliation{Shishan Laboratory, Nanjing University, Suzhou 215163, China}
\affiliation{Jiangsu Key Laboratory of Quantum Information Science and Technology, Nanjing University, Suzhou 215163, China}
\affiliation{Synergetic Innovation Center of Quantum Information and Quantum Physics, University of Science and Technology of China, Hefei, Anhui 230026, China}
\affiliation{Hefei National Laboratory, Hefei 230088, China}

\author{Xinsheng Tan}
\email{xstan@nju.edu.cn}
\affiliation{National Laboratory of Solid State Microstructures, School of Physics, Nanjing University, Nanjing 210093, China}
\affiliation{Shishan Laboratory, Nanjing University, Suzhou 215163, China}
\affiliation{Jiangsu Key Laboratory of Quantum Information Science and Technology, Nanjing University, Suzhou 215163, China}
\affiliation{Synergetic Innovation Center of Quantum Information and Quantum Physics, University of Science and Technology of China, Hefei, Anhui 230026, China}
\affiliation{Hefei National Laboratory, Hefei 230088, China}

\author{Yang Yu}
\email{yuyang@nju.edu.cn}
\affiliation{National Laboratory of Solid State Microstructures, School of Physics, Nanjing University, Nanjing 210093, China}
\affiliation{Shishan Laboratory, Nanjing University, Suzhou 215163, China}
\affiliation{Jiangsu Key Laboratory of Quantum Information Science and Technology, Nanjing University, Suzhou 215163, China}
\affiliation{Synergetic Innovation Center of Quantum Information and Quantum Physics, University of Science and Technology of China, Hefei, Anhui 230026, China}
\affiliation{Hefei National Laboratory, Hefei 230088, China}

\date{\today}
\maketitle

~\\
~\\
\newpage
\tableofcontents

~\\
~\\

The following Supplementary Material provides a detailed exposition of our methods. It is organized into three sections: the first presents the theoretical foundation of the butterfly metrology protocol; the second describes the sample properties and cryogenic circuit design; and the third outlines the key experimental implementations, including the out-of-time-ordered correlators (OTOCs) decomposition, Z-gate calibration, and time-reversed circuit construction.

\newpage

\section{Theoretical Foundation}
\subsection{Detailed analysis of the butterfly metrology protocol} \label{sec:1}
A generic quantum sensing protocol is structured in three stages: preparation of an initial state, interaction with the target signal to accumulate information, and finally, readout of a specific observable $M$. Our protocol operates within this framework but specifically leverages time-reversed dynamics---a feature shared by prominent techniques like echo protocols. However, a crucial distinction exists: in standard echo protocols, only a single cycle of forward and reverse evolution is used. The quantum state is prepared by a forward evolution from a product state, $U\ket{0}$, and the reverse evolution is applied only to assist in the final measurement. Consequently, such protocols can only achieve a quantum enhancement for a very specific class of unitary dynamics. 

While our "butterfly metrology" protocol fundamentally differs in its use of time reversal. Our protocol utilizes a combination of forward and reverse time evolution to prepare a "butterfly state", $\psi_B = (\ket{0}^{\otimes N}+ iV(t)\ket{0}^{\otimes N})/\sqrt{2}$, a superposition of a polarized state and a scrambled state with zero mean polarization, generated through forward evolution $U=e^{-i H t}$, a local perturbation $(I + i V)/\sqrt{2}$, and reverse evolution $(U^\dagger)$. 
As shown in the protocol circuit in Fig.~1 of the main text, measuring the local operator $V$ yields an expectation value given by:
\begin{equation}
\begin{aligned}
\langle V\rangle_\phi= & \frac{1}{2}\langle\mathbf{0}| V(t)|\mathbf{0}\rangle+\frac{1}{2}\langle\mathbf{0}| V(t) e^{i \phi S_z} V(t) e^{-i \phi S_z} V(t)|\mathbf{0}\rangle \\
& - \operatorname{Im}\left[e^{i \phi N / 2}\langle\mathbf{0}| V(t) e^{-i \phi S_z} V(t)|\mathbf{0}\rangle\right].
\end{aligned}
\end{equation}
For a small signal $\phi$, the first and second terms approach zero, leading to the approximation $\langle V \rangle_\phi \approx -\phi(N/2-\bra{0} V(t) S_z V(t) \ket{0})$. The sensitivity at $\phi=0$ is therefore:
\begin{equation}
    \frac{1}{\eta_{\phi=0}} = \frac{N}{2} - \bra{0} V(t) S_z V(t) \ket{0}.
\end{equation}
Using the identity $\sigma_i^z \ket{0} = 1$, this expression can be rewritten as:
\begin{equation}
    \frac{1}{\eta_{\phi=0}} = \frac{1}{2} \sum_i (1-\bra{0} \sigma_i^z V(t) \sigma_i^z V(t) \ket{0}),
\end{equation}
Here, each expectation value corresponds to a local out-of-time-order correlator (OTOC).

These OTOCs quantify whether the time-evolved perturbation, $V(t)$, commutes with each spin operator $\sigma^z_i$. 
As a function of time, each OTOC begins at a value of unity and decays to zero as the operator $V(t)$ grows to have support on qubit $i$. The sum in Equation (3) thus counts the number of qubits $N$, within the support of $V(t)$. Intuitively, as $V(t)$ scrambles each qubit within its support, it effectively randomizes the qubit's polarization. This process creates a polarization difference of approximately $N/2$ between the scrambled state, $V(t) \ket{0}$, and the original polarized state, $\ket{0}$. Consequently, this yields a sensitivity of $\eta \approx 2/N$.

We further undertake a novel analysis of the local sensitivity. As established in the preceding equation, the expectation values of the initial two terms are zero, leaving the final term as the sole contributor to the signal. To evaluate this term, we expand the perturbed state in the computational basis: $V(t) \ket{0} = \sum_{s \in \{0,1\}^{\otimes N}} C_s \ket{s}$. Central to our analysis is the polarization distribution, defined as $P(S_z) = \sum_{|s|=2S_z} |C_s|^2$, where $|s|=\sum_i (-1)^{s_i}$ and $S_z = -N/2, -N/2+1,..., N/2-1, N/2$. Define the characterisitic sum
\begin{equation}
    \chi(\phi) = \sum_{Sz} e^{-i\phi Sz} P(Sz),
\end{equation}
then the expectation value takes the form of
\begin{equation}
    \langle V\rangle_\phi= -Im[e^{i \phi N/2} \chi(\phi)].
\end{equation}
We Taylor expand this equation around $\phi=0$ up to third order:
\begin{equation}
\begin{gathered}
e^{i \phi N / 2}=1+i \frac{N}{2} \phi-\frac{(N / 2)^2}{2} \phi^2-i \frac{(N / 2)^3}{6} \phi^3+\mathcal{O}\left(\phi^4\right) \\
\chi(\phi)=1-i \mu \phi-\frac{m_2}{2} \phi^2+i \frac{m_3}{6} \phi^3+\mathcal{O}\left(\phi^4\right),
\end{gathered}
\end{equation}
where we define the moments $\mu=\left\langle S_z\right\rangle=\sum_{S_z} S_z P\left(S_z\right), \quad m_2=\left\langle S_z^2\right\rangle, \quad m_3=\left\langle S_z^3\right\rangle$.
Multiply and take the imaginary part. The first nonzero terms are
\begin{equation}
\langle V\rangle_\phi=-\phi\left(\frac{N}{2}-\mu\right)-\phi^3\left(\frac{m_3}{6}-\frac{N m_2}{4}+\frac{\mu N^2}{8}-\frac{N^3}{48}\right)-\mathcal{O}\left(\phi^5\right) .
\end{equation}
Keep the first order term, we have
\begin{equation}
    \langle V\rangle_\phi \approx -\phi\left(\frac{N}{2}-\mu\right).
\end{equation}

\subsection{Loschmidt echo}
Loschmidt echo measurements validate the resonant tuning essential to our quantum sensing protocol.
As shown in Fig.~2a of the main text, the echo circuit includes initial state 
preparation $R_y^0(\pi/2)|g\rangle_0$, followed by the Loschmidt echo sequence 
$U(-t)U(t)$ applied to the complete lattice, and concluded with a tomography 
pulse $R_y^0(\pi/2)$ before readout. The complete sequence implements
\begin{equation}
    R_y^0(\pi/2) \; U(-t)U(t) \; R_y^0(\pi/2) |g\rangle_0 \otimes |\psi\rangle = |e\rangle_0 \otimes |\psi\rangle,
\end{equation}
where $|\psi\rangle$ denotes the state of the remaining eight qubits.

Measured deviations from this 
ideal behavior quantify time-reversal fidelity. For perfect time reversal with $U(t)U(-t)=I$, qubit 0 should remain excited. However, we observe a gradual decay of exciation probability with increasing Loschmidt echo time $t$, consistent with qubit dephasing during evolution.


\subsection{Circuits for the OTOC experiment: OTOC decomposition}
Here we demonstrate the experimental process of the OTOC experiment defined in Eq.~(5) of the main text.
We take $V=\sigma_i^x$, and $W=\sigma_j^z$, with $F(t)$ takes the form of 
\begin{equation}
    F(t) = \bra{\psi} \sigma_j^z(t) \sigma_i^x \sigma_j^z(t) \sigma_i^x \ket{\psi}.
\end{equation}
The OTOC $F(t)$ is not a time-ordered correlator, preventing its direct measurement via local operators. Despite this, the target of our measurement is the Hermitian operator $C$ from Eq.~(6) of the main text, which can be accessed experimentally
\begin{equation}
    C = 2 - \bra{\psi} \sigma_j^z(t) \sigma_i^x \sigma_j^z(t) \sigma_i^x + \sigma_i^x \sigma_j^z(t) \sigma_i^x \sigma_j^z(t) \ket{\psi}.
\end{equation}
The sum above is challenging for direct gate-based implementation, as it necessitates a complete and frequently intractable diagonalization of the Hamiltonian. To circumvent this impediment and facilitate the experimental determination of the squared commutator $C$, we invoke a mathematical equivalence obtained by expanding
\begin{equation}
\begin{aligned}
&\langle\psi| \hat{\sigma}_j^z(t) \hat{\sigma}_i^x \hat{\sigma}_j^z(t) \hat{\sigma}_i^x|\psi\rangle=\langle\psi| \hat{\sigma}_j^z(t) \hat{\sigma}_i^x \hat{\sigma}_j^z(t) \frac{1+\hat{\sigma}_i^x}{2}|\psi\rangle-\langle\psi| \hat{\sigma}_j^z(t) \hat{\sigma}_i^x \hat{\sigma}_j^z(t) \frac{1-\hat{\sigma}_i^x}{2}|\psi\rangle \\
&=\langle\psi| \frac{1+\hat{\sigma}_i^x}{2} \hat{\sigma}_j^z(t) \hat{\sigma}_i^x \hat{\sigma}_j^z(t) \frac{1+\hat{\sigma}_i^x}{2}|\psi\rangle-\langle\psi| \frac{1-\hat{\sigma}_i^x}{2} \hat{\sigma}_j^z(t) \hat{\sigma}_i^x \hat{\sigma}_j^z(t) \frac{1-\hat{\sigma}_i^x}{2}|\psi\rangle \\
&+\langle\psi| \frac{1-\hat{\sigma}_i^x}{2} \hat{\sigma}_j^z(t) \hat{\sigma}_i^x \hat{\sigma}_j^z(t) \frac{1+\hat{\sigma}_i^x}{2}|\psi\rangle-\langle\psi| \frac{1+\hat{\sigma}_i^x}{2} \hat{\sigma}_j^z(t) \hat{\sigma}_i^x \hat{\sigma}_j^z(t) \frac{1-\hat{\sigma}_i^x}{2}|\psi\rangle, \\
&\langle\psi| \hat{\sigma}_i^x \hat{\sigma}_j^z(t) \hat{\sigma}_i^x \hat{\sigma}_j^z(t)|\psi\rangle=\langle\psi| \frac{1+\hat{\sigma}_i^x}{2} \hat{\sigma}_j^z(t) \hat{\sigma}_i^x \hat{\sigma}_j^z(t) \frac{1+\hat{\sigma}_i^x}{2}|\psi\rangle-\langle\psi| \frac{1-\hat{\sigma}_i^x}{2} \hat{\sigma}_j^z(t) \hat{\sigma}_i^x \hat{\sigma}_j^z(t) \frac{1-\hat{\sigma}_i^x}{2}|\psi\rangle \\
&+\langle\psi| \frac{1+\hat{\sigma}_i^x}{2} \hat{\sigma}_j^z(t) \hat{\sigma}_i^x \hat{\sigma}_j^z(t) \frac{1-\hat{\sigma}_i^x}{2}|\psi\rangle-\langle\psi| \frac{1-\hat{\sigma}_i^x}{2} \hat{\sigma}_j^z(t) \hat{\sigma}_i^x \hat{\sigma}_j^z(t) \frac{1+\hat{\sigma}_i^x}{2}|\psi\rangle, \\
& \Rightarrow C=2-\langle\psi| \hat{\sigma}_j^z(t) \hat{\sigma}_i^x \hat{\sigma}_j^z(t) \hat{\sigma}_i^x|\psi\rangle-\langle\psi| \hat{\sigma}_i^x \hat{\sigma}_j^z(t) \hat{\sigma}_i^x \hat{\sigma}_j^z(t)|\psi\rangle= \\
& 2-2\left[\langle\psi| \frac{1+\hat{\sigma}_i^x}{2} \hat{\sigma}_j^z(t) \hat{\sigma}_i^x \hat{\sigma}_j^z(t) \frac{1+\hat{\sigma}_i^x}{2}|\psi\rangle-\langle\psi| \frac{1-\hat{\sigma}_i^x}{2} \hat{\sigma}_j^z(t) \hat{\sigma}_i^x \hat{\sigma}_j^z(t) \frac{1-\hat{\sigma}_i^x}{2}|\psi\rangle\right] .
\end{aligned}
\end{equation}
It is evident that
\begin{equation}
    C = 2 + C_- - C_+
\end{equation}
where 
\begin{equation}
C_{ \pm} \equiv\langle\psi| \frac{1 \pm \hat{\sigma}_i^x}{\sqrt{2}} \hat{\sigma}_j^z(t) \hat{\sigma}_i^x \hat{\sigma}_j^z(t) \frac{1 \pm \hat{\sigma}_i^x}{\sqrt{2}}|\psi\rangle.
\end{equation}
The new quantities $C_{\pm}$ are experimental observables, as they are both Hermitian and possessing the symmetric form.

\section{Experimental setup and system characterization}

\subsection{Experimental Setup}

\begin{figure*}
		\begin{minipage}[b]{1.0\textwidth}
			\centering
			\includegraphics[width=13cm]{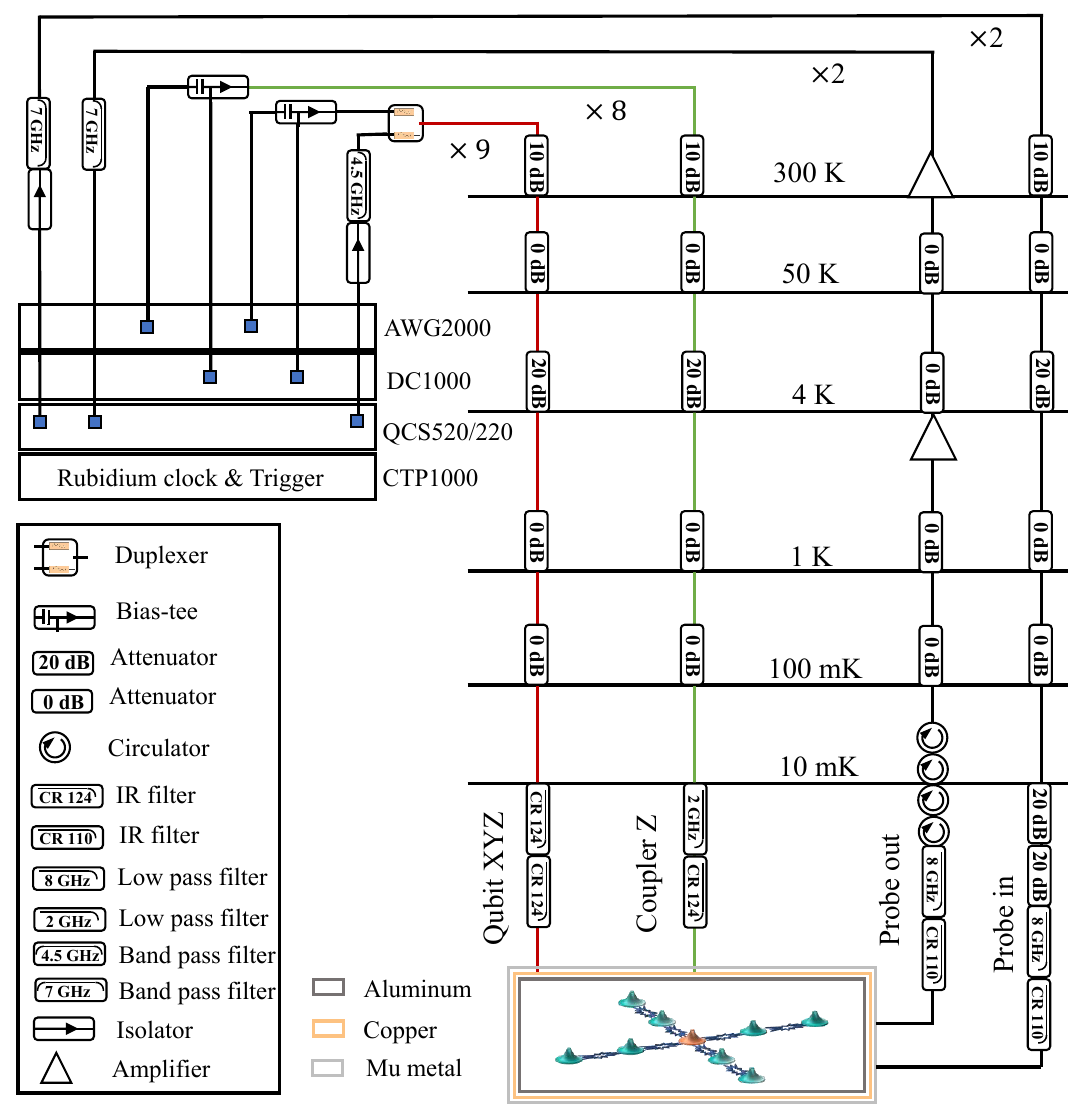}
		\end{minipage}
		\caption{\justifying 
			\textbf{Wiring diagram.}
			The microwave pulses for XY control and probe are produced by up-conversion technique, 
			which synthesize a microwave carrier from a local oscillator is modulated 
			by intermediate frequency (IF) signal from an arbitrary wave generator (for XY control pulse, it is Tektronix-5208, while Keysight M3202A for probe signal), using an IQ mixer. 
			For Z control, it is a static flux bias  signal which is directly generated by Tektronix-5208. 
			In oder to achieve cryogenic quantum control and probe, all signals are attenuated and filtered before sent into the quantum chip through corresponding lines.
			For readout, the output signal passes through two circulators and is then filtered 
			after the probe signal with quantum states information comes out of the chip. 
			Then it is amplified by a high-electron-mobility-transistor (HEMT) at low temperature.
			and is amplified again after coming out of the fridge to increase signal noise ratio (SNR).
			Before digital processing by a ADC channel (ATS-9870), the readout signal is down converted to IF signal 
			and the another channel of ATS-9870 board received a IF signal from Keysight M3202A to provide a reference phase.
			\label{fig:FIGS1}
			}
\end{figure*}

Our experiment is performed on a 2D cross-shaped lattice of coupler-coupled transmon qubits~\cite{Barends2014,Yan2018,Arute2019,Sung2021}. The couplers have, facilitating a tunable coupling range between -2.5 MHz to 10 MHz.
Each qubit is coupled to an XYZ driving line used for both XY control and flux control (with a max bandwith of 200 MHz). 

The superconducting quantum chip used in this work contains nine transmon qubits, each consisting of a cross-shaped shunting capacitance and two Josephson junctions forming a SQUID loop~\cite{Barends2013}. 
Each qubit is inductively coupled to an XYZ control line for modulating its transition frequency, and dispersively coupled to an individual readout resonator, which in turn is coupled to a common transmission line on the other side.
The detailed parameters for all nine qubits, including frequencies, anharmonicities, coherence times, and cavity frequencies, 
are summarized in Table~\ref{table:qubit_params}. 
All the room and cryogenic instruments and circuits used to control and readout are shown in Fig.~\ref{fig:FIGS1}.
As shown in Fig.~\ref{fig:FIGS1}, flux control is used to tune the transmon frequency and bias it at the desired operating point, XY driving is employed to set the Rabi frequency and phase for transitions between transmon energy levels, and the probe line is used to acquire the readout signal.

Our measurements are performed in a dilution refrigerator (BlueFors LD-400) with a base temperature of approximately $10~\mathrm{mK}$. In this work, all attenuators and filters installed in the cryogenic microwave lines are supplied by HERMERCS SYSTEM.
These components are employed to suppress excess thermal photons originating from higher-temperature stages of the dilution refrigerator~\cite{Krinner2019,Danilin2022}.
Similarly, the readout signals are measured through a dedicated microwave line, which includes two cryogenic isolators placed after the sample to prevent the qubit states from being affected by back-action noise from the readout line~\cite{Huang2021,Roy2015}.
The readout signal is amplified by 30~dB using a high-electron-mobility transitor (HEMT) provided by ZWDX, which is thermally anchored to the 4~K stage of the dilution refrigerator.
The signal is further amplified by approximately 70~dB at room temperature using a low-nosie amplifier (HWF0408-75-15) provided by Hengwei Microwave, and is finally sent to an ADC module, which is also manufactured by ZWDX.
Additionally, the parameterized pulses for XY control and probe signals are programmed in Python and directly generated by a QCS520 integrated microwave measurement and control system (14 bits, 1~GHz sampling rate).
All components for generating microwave signals are frequency locked by Rubidium atomic clock (CTP1000) with 200 MHz made by ZWDX.
 
\begin{table}
  \centering
  \caption{Qubit parameters}
    \label{table:qubit_params}
  \begin{tabular}{c *{9}{c}}
    \toprule
    \hline
    \textbf{Qubit} & $\mathbf{Q_0}$ & $\mathbf{Q_1}$ & $\mathbf{Q_2}$ & $\mathbf{Q_3}$ & $\mathbf{Q_4}$ & $\mathbf{Q_5}$ & $\mathbf{Q_6}$ & $\mathbf{Q_7}$ & $\mathbf{Q_8}$ \\
    \midrule
	$\omega_r/2\pi$ (GHz) & 6.8548 & 6.7988 &  6.9037 & 6.8216 & 6.8763 & 6.7433 & 6.9473 &  6.7640 & 6.9206 \\ \hline
	$\omega_q^{sp}/2\pi$ (GHz) & 4.9565 & 4.9125 &  4.6940 & 4.7660 & 4.8097 & 4.8925 & 4.8411 &  4.8479 & 4.7810 \\ \hline
 	$\alpha/2\pi$ (MHz) & -244 & -234 &  -248 & -246 & -265 & -244 & -248 &  -264 & -262 \\ \hline
	$T_1^{sp}$ ($\mu$s) & 58.06 & 47.58 &  75.62 & 66.98 & 52.80 & 24.76 & 55.68 &  74.28 & 70.25 \\ \hline
	$T_2^{sp}$ ($\mu$s) & 10.00 & 9.46 &  20.00 & 13.81 & 11.41 & 15.96 & 56.51 &  45.17 & 44.02 \\ \hline
    $T_2^{work}$ ($\mu$s) & 1.77 & 2.08 &  2.70 & 2.91 & 2.54 & 1.88 & 1.99 &  1.99 & 2.03 \\ \hline
    \bottomrule
  \end{tabular}
\end{table}

\subsection{System characterization}

\subsubsection{Resonant evolution and frequency compensation}
The forward unitary evolution $U(t)$, a fundamental element of our metrology protocol, is executed by bringing all nine qubits into mutual resonance. This resonant configuration is maintained with an effective nearest-neighbor coupling strength of $g_{eff} = 5~\mathrm{MHz}$, achieved by fixing the tunable couplers at predetermined frequencies. We actively compensate frequency shifts arising from coupler-qubit interactions to ensure precise resonance throughout the evolution interval $t$.
To  calibrate the effective coupling strength ($g_{eff}$) between two nearest-neighbour qubits, we bring the qubits into resonance and vary the flux bias applied to their intermediate coupler~\cite{Sung2021,Yan2018,Collodo2020,Xu2020}.
The coupling strength $g_{eff}$ is extracted by fitting the vacuum Rabi oscillations measured at different coupler flux points, where the oscillation frequency corresponds to $2|g_{eff}|$. 

When the tunable couplers are biased to their operating points ($g_{eff}\approx 5~\mathrm{MHz}$ in our device), the strong XX-YY type interaction between the couplers and qubits introduces additional qubit frequency shifts due to static dispersive coupling. As the coupler frequency approaches that of the qubits, the effective qubit energies are renormalized, leading to detuning  that must be precisely compensated to maintain qubit resonace during the forward evolution. In our architecture, each adjacent coupler at its operating point induces a qubit frequency shift of approximately $5-6~\mathrm{MHz}$.
To counteract this effect, we bias other qubits to $\omega_-$ = $\omega_{res}-\Delta$.(In our case, a large detuning $\Delta/2\pi$ = 100MHz ($\Delta$ = $20 g_{eff}$) is used to effectively isolate the target qubit from the others.) Concurrently, we detune each coupler to the operating point using flux pulse, and then apply a calibrated DC-flux bias to correct the target qubit frequency to resonance ($\omega_{res}/2\pi$ = 4.6GHz).

The compensation procedure is essential for maintaining precise resonant exchange interactions and suppressing residual $Z$-type couplings, which if uncompensated, would perturb coherent dynamics and reduce metrological sensitivity.


\subsubsection{Realization and calibration of the time-reversed unitary evolution U(-t)}\label{subsec:II B}

\begin{figure*}
		\begin{minipage}[b]{1.0\textwidth}
			\centering
			\includegraphics[width=0.95\textwidth]{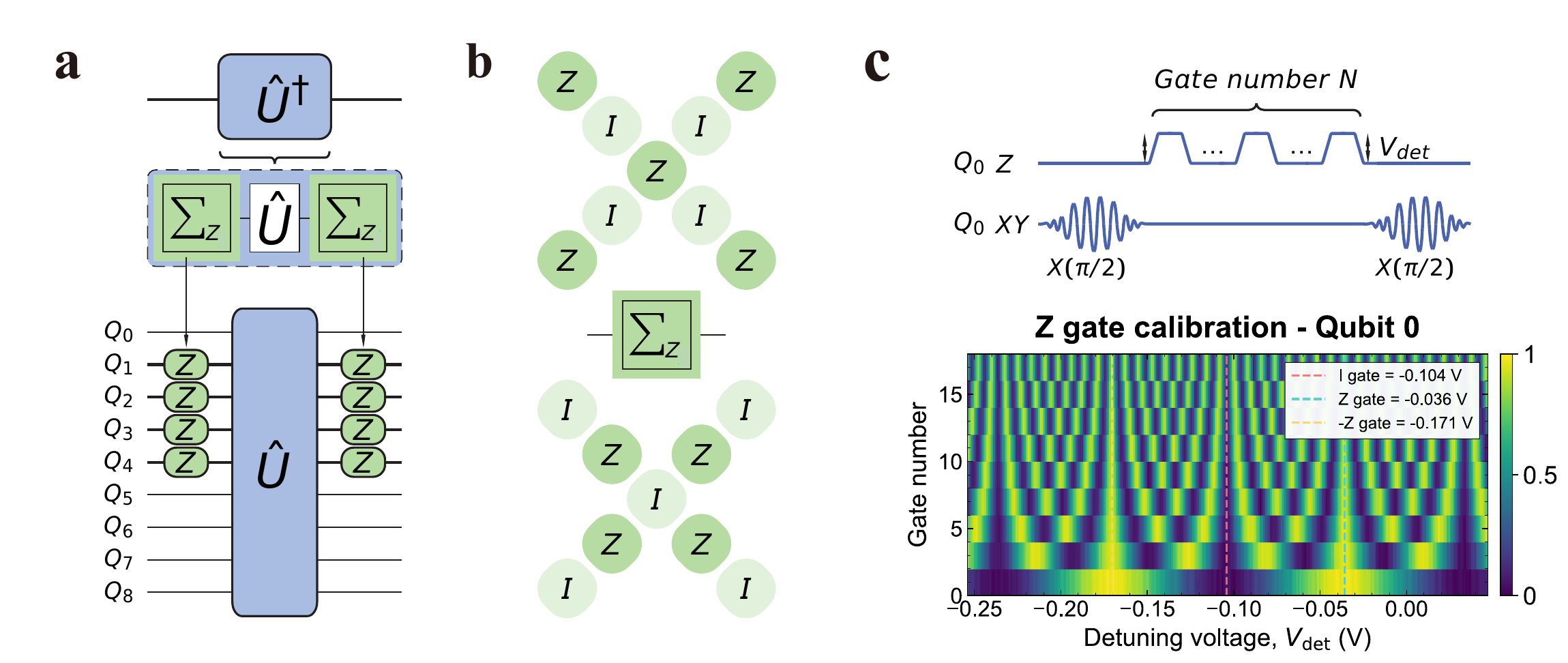}
		\end{minipage}
        \caption{\textbf{Implementation and calibration of the time-reversed evolution \(U(-t)\).} 
\textbf{a,} Quantum circuit for generating \(U(-t)\) by conjugating the forward evolution \(U(t)\) with Pauli-Z gates (\(\Sigma_z\)).
\textbf{b,} Two possible configurations of Pauli-Z gates on the 9-qubit cross geometry that realize the sign inversion of the Bose--Hubbard Hamiltonian. The experimentally chosen configuration applies Z gates to the four first-neighbor qubits (q1--q4, \textit{lower} schematic). An alternative configuration applies Z gates to the central qubit (q0) and the outer qubits (q5--q8, \textit{upper} schematic). Both satisfy \(\Sigma_z H \Sigma_z = -H\) for nearest-neighbor hopping.
\textbf{c,} Calibration of the flux-driven \(Z(\pi)\) gate. \textit{Upper:} Ramsey-type pulse sequence embedding an odd number \(N\) of \(Z\) pulses with amplitudes \(\Delta A_i\). The phase accumulated per pulse is extracted from Ramsey fringe oscillations. \textit{Lower:} Extracted phase shift as a function of \(N\), converging to \(\pi\),\(-\pi\) and \(0\) as the number of gates increases, confirming a self-consistent calibration of the \(Z(\pi)\), \(Z(-\pi)\) and \(I\) gates.}
			\label{fig:FIGS2}
\end{figure*}

In our experiment, we implement reverse time evolution by constructing a Hamiltonian with an inverted sign, a method similarly suggested for experimental access to the OTOC. This is realized by sandwiching the forward time evolution $U(t)$ under Hamiltonian $H$ between two identical sets of single-qubit Pauli-Z gates. These Pauli-Z gates, denoted as $\Sigma_z = \Pi_i \sigma_i^Z$, are applied to every other qubit in the lattice.
Fig.~\ref{fig:FIGS2}a shows the quantum circuit used to realize the reverse evolution \(U(-t)\) on our 9-qubit superconducting processor. In the experiment, Pauli-Z gates are applied to the four qubits directly neighboring the central qubit, q0 (labeled q1--q4). Fig.~\ref{fig:FIGS2}b illustrates the two distinct Z-gate configurations compatible with the cross geometry of the chip: the experimentally implemented configuration, with Z gates on the first-neighbor qubits q1--q4 (lower schematic), and an alternative configuration, with Z gates on the central qubit q0 and the four outer qubits q5--q8 (upper schematic).

In the Bose-Hubbard model, we find that $\sum_z a_i^\dagger a_j \sum_z = - a_i^\dagger a_j$ for any pair of adjacent qubits $\langle i,j \rangle$. Consequently, 
\begin{equation}
    \sum\nolimits_{z} H \sum\nolimits_{z} = \sum_{\langle i,j \rangle} -J_{ij} a_i^\dagger a_j = -H  
\end{equation}
The natural time evolution under this transformation satisfies
\begin{equation}
    \sum\nolimits_{z} U(t)\sum\nolimits_{z} = \sum\nolimits_{z} e^{-i H t} \sum\nolimits_{z} = e^{-i(-H)t} = U(-t),
\end{equation}
thus realizing the desired time-reversed evolution.

As described in \hyperref[subsec:II B]{section II B}, the implementation of the time-reversed unitary evolution $U(-t)$ requires precise calibration of the flux-driven $Z(\pi)$ gate~\cite{Rol2019,Werninghaus2021,Wittler2021,McKay2017}. We realize the Z-rotation through a flux-pulse-induced detuning of the qubit frequency, which accumulates a controlled dynamic phase. To calibrate the pulse amplitude corresponding to a $\pi$-phase rotation, we embed an odd number $N$ of $Z$-pulses with individual amplitudes $\Delta A_i$ into a Ramsey-type interferometric sequence, as illustrated in Fig.~\ref{fig:FIGS2}c. The accumulated phase shift is extracted from the Ramsey fringe oscillations as a function of $N$, allowing us to precisely determine the effective phase per pulse. As shown in Fig.~\ref{fig:FIGS2}c, the extracted phase converges toward a stable $\pi$ value with increasing gate count, confirming the self-consistency of the calibration.

This iterative calibration procedure effectively compensates for residual nonlinearites in the flux response and slow drifts in the qubit frequency. Once the $Z(\pi)$ pulse amplitude is determined, it is used as a reference to define the flux-pulse amplitude-to-phase conversion curve for arbitrary $Z(\theta)$ rotations. The calibrated $Z$-gates exhibit high phase fidelity, ensuring accurate realization of the reversed evolution $U(-t)$ required in the Loschmidt-echo and butterfly metrology protocols.

\begin{figure*}
		\begin{minipage}[b]{1.0\textwidth}
			\centering
            \includegraphics[width=0.8\textwidth]{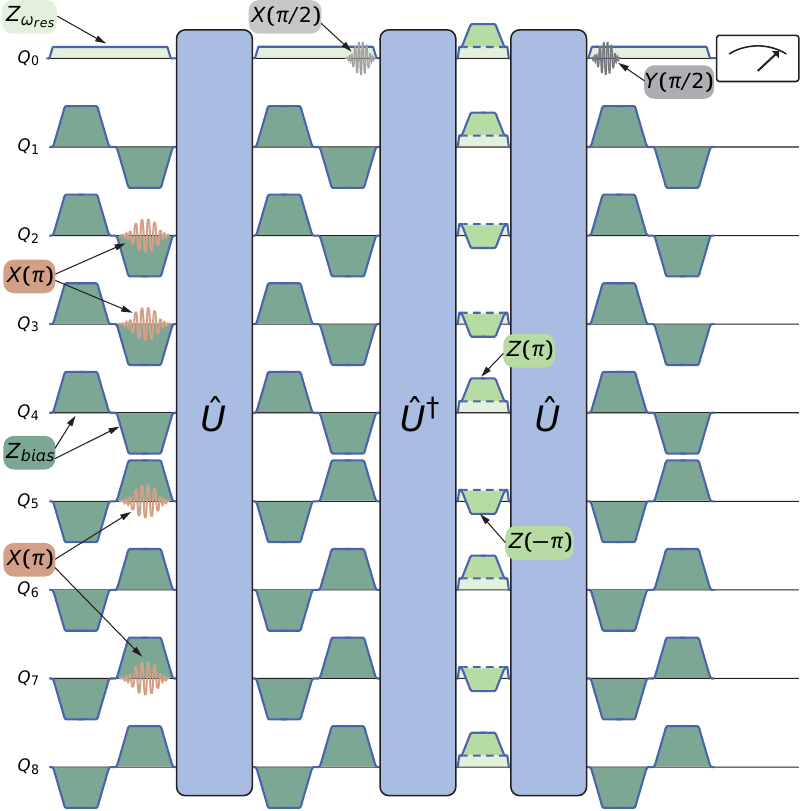}
		\end{minipage}
		\caption{\justifying 
			{\textbf{Detailed pulse sequence for the butterfly metrology protocol.} The complete experimental timeline is shown for all nine qubits (q0--q8). Microwave (XY) pulses for single-qubit gates are shown as brown colored waveforms, while net-zero Z pulses for frequency tuning and phase correction are shown as green blocks. The protocol comprises three stages: state preparation, where the perturbation $I+iV$ is applied; signal sensing, where qubits are detuned in a checkerboard pattern to freeze intrinsic dynamics, except for the central qubit q0, which remains at the fixed reference frequency $\omega_q = 4.6\ \text{GHz}$; and final readout, where only the central qubit (q0) is measured.}
			\label{fig:FIGS3}
			}
\end{figure*}

\subsubsection{Detailed pulse sequence for butterfly metrology protocol}
The butterfly metrology protocol follows the three-stage circuit shown in Fig.~1c of the main text: butterfly state preparation, signal sensing, and final readout. In the preparation stage, a perturbation operator \(I+iV\) disrupts the perfect cancellation between the forward evolution \(U\) and its reverse \(U^{\dagger}\), generating a scrambled state that is subsequently interfered with a simple polarized state \(\ket{0}^{\otimes N}\). 

A complete and detailed implementation of this protocol on the 9-qubit processor is provided in Fig.~\ref{fig:FIGS3}. To maximally freeze the intrinsic system dynamics during the sensing stage and thus extend the coherence time of the scrambled state, qubits are detuned according to a checkerboard pattern. This ensures that adjacent sites are shifted in opposite frequency directions relative to their idle frequencies. The central qubit (q0) constitutes the sole exception; it remains fixed at the resonant frequency \(\omega_q = 4.6\ \text{GHz}\) (in the rotating frame) throughout the entire sequence to serve as a stable phase reference. For all initial state preparations, qubit operations, and final readout, we employ calibrated microwave pulses together with net-zero Z control pulses to ensure accurate phase and timing alignment across the device.

\subsubsection{Flux distortion}
\begin{figure*}
		\begin{minipage}[b]{1.0\textwidth}
			\centering
            \includegraphics[width=0.90\textwidth]{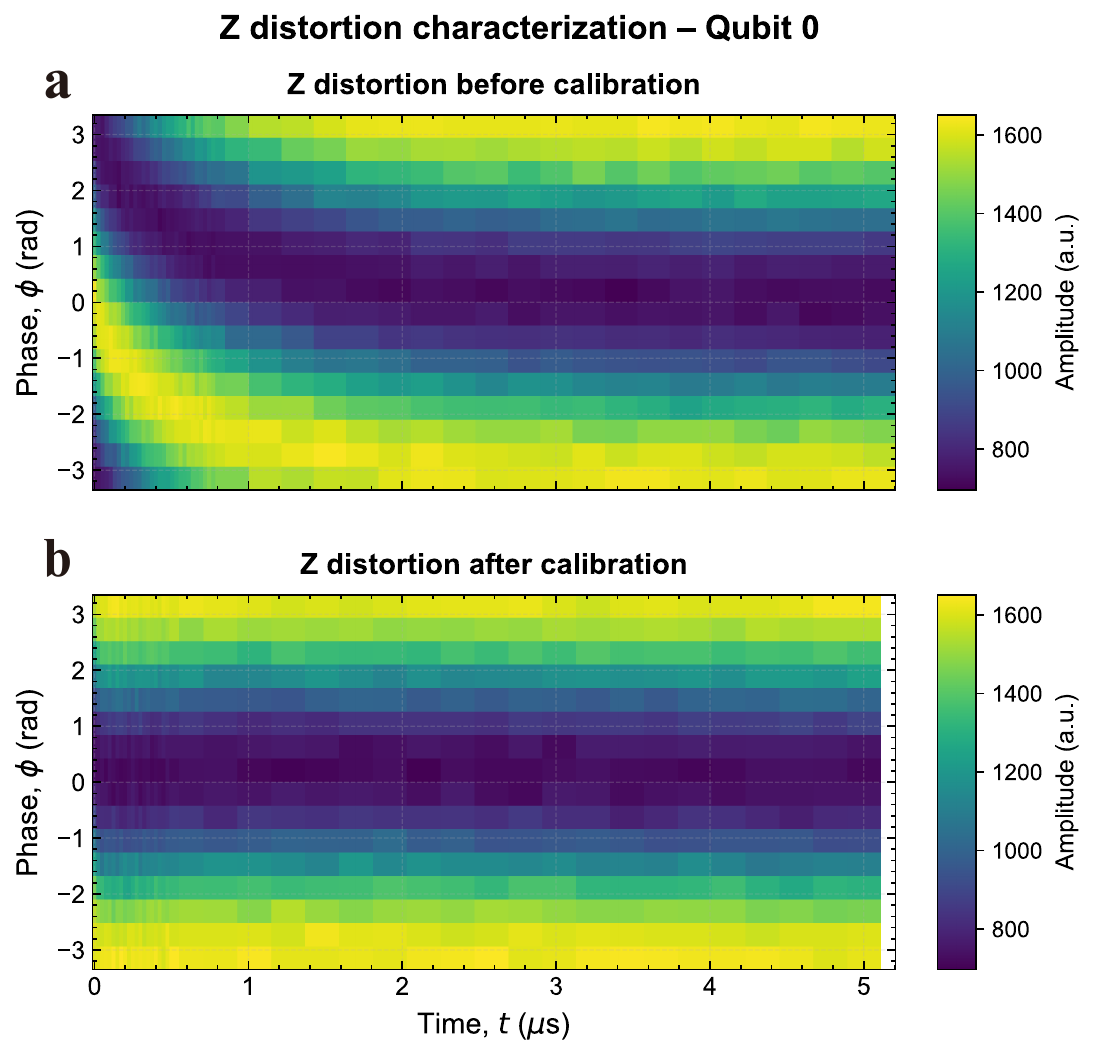}
		\end{minipage}
   \caption{\justifying
\textbf{Characterization and calibration of flux pulse temporal distortions.}
Amplitude response of transmon qubit Q0 measured as a function of flux pulse duration and accumulated phase during flux pulse execution.
\textbf{a,} Distortion profile measured before calibration, revealing temporal deviations induced by waveform distortions along the flux control line.
\textbf{b,} Distortion profile after applying predistortion calibration, demonstrating a substantial suppression of temporal distortions and a more faithful realization of the intended flux pulse waveform.
}
	\label{fig:FIGS4}
\end{figure*}

As flux pulses propagate from the room-temperature electronics to the sample through various electrical components inside the dilution refrigerator, they inevitably experience waveform distortions~\cite{Rol2017,Gustavsson2016,Wittler2021}. These distortions arise from impedance mismatches, finite bandwidth of filters and attenuators, and reflections along the transmission line, all of which can significantly degrade the accuracy of flux-based control operations. In practice, such effects manifest as exponential overshoot or undershoot features in the step response of the flux control line, leading to temporal deviations between the intended and the actual flux bias experienced by the qubit.

Accurate characterization and compensation of these distortions are therefore essential for achieving high-fidelity gate calibration and reliable implementation of flux-tunable operations. A commonly used phenomenological model for the step response can be written as
\begin{equation}
V_{\mathrm{qubit}}(t) = \left(1 + A e^{-t/\tau}\right) V_{\mathrm{AWG}}(t),
\end{equation}
where $V_{\mathrm{qubit}}(t)$ is the effective control signal perceived by the qubit, $V_{\mathrm{AWG}}(t)$ is the waveform generated by the arbitrary waveform generator (AWG), $\tau$ is the characteristic time constant of the distortion, and $A$ denotes its amplitude.

To experimentally characterize the temporal distortion, we employ the qubit itself as a sensitive probe using Ramsey interferometry (Fig.~\ref{fig:FIGS4}a). During the flux pulse, distortions induce a time-dependent detuning $\delta(t)$, which accumulates as an additional phase shift $\phi(t) = \int_0^t \delta(t)\,\mathrm{d}t$. By fitting the measured phase evolution $\phi(t)$, we extract the parameters $A$ and $\tau$, thereby reconstructing the effective step response of the flux line.

Based on the extracted distortion model, we subsequently apply a predistortion correction to the AWG waveform to compensate for the measured temporal response. The effectiveness of this calibration procedure is demonstrated in Fig.~\ref{fig:FIGS4}b, where the distortion profile is substantially suppressed and the accumulated phase exhibits a markedly improved agreement with the ideal flux pulse shape. This calibration enables a more accurate realization of flux control pulses and forms the basis for the high-fidelity flux-tunable operations used throughout this work.

\subsubsection{Calibration of Flux Noise from Qubit Dephasing Measurements}

A dominant source of dephasing in superconducting qubits is low-frequency flux noise, which causes random shifts in the qubit's transition frequency~\cite{Kakuyanagi2007,Yan2012,Kumar2016,Rower2023}. This section details the theoretical model and calibration procedure used to quantify the amplitude of this flux noise, $\sigma_\Phi$, by measuring the flux dependence of the qubit's pure dephasing rate, $\Gamma_\phi$.

We model the qubit frequency $f$ as a stochastic variable following a Gaussian (normal) distribution around a mean frequency $f_0$, with a standard deviation $\sigma_f$~\cite{Yan2012,Yan2016}:
\begin{equation}
f \sim \mathcal{N}(f_0, \sigma_f^2)
\end{equation}
The corresponding probability density function is:
\begin{equation}
y(f) = \frac{1}{\sigma_f \sqrt{2\pi}} \exp\left(-\frac{(f - f_0)^2}{2\sigma_f^2}\right)
\end{equation}

To find the expected dephasing, we consider a Ramsey experiment. A qubit initialized in the $|+\rangle$ state will accumulate a random phase $\phi(t) = 2\pi \int f(t') dt'$. For quasi-static frequency noise (valid over the timescale of the experiment)~\cite{Yan2012,Yan2016}, the expectation value of the X-projection, $\langle \sigma_x \rangle$, decays as:
\begin{equation}
\langle \sigma_x \rangle(t) = \int \cos(2\pi f t)  y(f)  df
\end{equation}
Substituting $f = f_0 + x$, where $x$ is the frequency fluctuation, the integral becomes:
\begin{equation}
\langle \sigma_x \rangle(t) = \cos(2\pi f_0 t) \cdot \int \cos(2\pi x t) \frac{1}{\sigma_f \sqrt{2\pi}} e^{-\frac{x^2}{2\sigma_f^2}} dx
\end{equation}
This integral is the Fourier transform of a Gaussian, yielding:
\begin{equation}
\langle \sigma_x \rangle(t) = \cos(2\pi f_0 t) \cdot \exp\left(-\frac{(2\pi \sigma_f)^2 t^2}{2}\right)
\end{equation}
The decay envelope is thus a Gaussian function, $\exp(-\Gamma_{\phi}^2 t^2)$, allowing us to identify the dephasing rate due to frequency noise as:
\begin{equation}
\Gamma_{\phi} = \frac{2\pi \sigma_f}{\sqrt{2}}
\label{eq:gamma_f}
\end{equation}

The qubit frequency is tunable via an external magnetic flux $\Phi$. Near the maximum frequency (the ``flux sweet spot'' or ``top of the spectrum''), the dispersion relation can be approximated by an inverted parabola~\cite{Yan2012,Rower2023}:
\begin{equation}
f(\Phi) = -a \Phi^2 + c
\end{equation}
where $a$ and $c$ are positive constants. The slope, or the flux sensitivity, is:
\begin{equation}
\frac{\partial f}{\partial \Phi} = -2a\Phi
\end{equation}
Defining $A = 2a$ (a positive constant representing the curvature), the magnitude of the sensitivity is:
\begin{equation}
\left| \frac{\partial f}{\partial \Phi} \right| = A |\Phi|
\label{eq:sensitivity}
\end{equation}
For small flux fluctuations $\sigma_\Phi$, the resulting standard deviation of the frequency noise is:
\begin{equation}
\sigma_f = \left| \frac{\partial f}{\partial \Phi} \right| \sigma_{\Phi} = A |\Phi| \sigma_{\Phi}
\label{eq:sigma_f}
\end{equation}

Substituting Eq.~\eqref{eq:sigma_f} into Eq.~\eqref{eq:gamma_f} gives the contribution of flux noise to the dephasing rate:
\begin{equation}
\Gamma_{c,\phi} = \frac{2\pi}{\sqrt{2}} A |\Phi| \sigma_{\Phi}
\end{equation}
For clarity, we can define a composite parameter $k$:
\begin{equation}
\Gamma_{c,\phi} = k |\Phi|, \quad \text{where} \quad k = \frac{2\pi}{\sqrt{2}} A \sigma_{\Phi}
\label{eq:k_param}
\end{equation}

The total measured dephasing rate $\Gamma_{\phi}$ has contributions from both this flux-dependent noise and a flux-independent intrinsic noise $\Gamma_{i,\phi}$~\cite{Kakuyanagi2007,Yan2012} (e.g., from charge noise or measurement noise). The combined dephasing rate is the quadrature sum:
\begin{equation}
\Gamma_{\phi} = \sqrt{\Gamma_{i,\phi}^2 + \Gamma_{c,\phi}^2} = \sqrt{\Gamma_{i,\phi}^2 + (k |\Phi|)^2}
\label{eq:total_gamma}
\end{equation}

The calibration procedure is as follows:
\begin{enumerate}
    \item \textbf{Measure:} Perform Ramsey experiments at a series of flux bias points, $\Phi$, detuned from the sweet spot.
    \item \textbf{Fit:} At each point, fit the decay of the Ramsey fringes to a Gaussian envelope, $\exp(-\Gamma_{\phi}^2 t^2)$, to extract the total dephasing rate $\Gamma_{\phi}$.
    \item \textbf{Model Fit:} Plot the measured $\Gamma_{\phi}$ as a function of $|\Phi|$ and fit the data to the model in Eq.~\eqref{eq:total_gamma}. This fit yields two parameters: the intrinsic dephasing rate $\Gamma_{i,\phi}$ and the crucial slope parameter $k$.
\end{enumerate}


\begin{table}
  \centering
  \caption{Calibrated Flux Noise Parameters}
  \label{table:flux_noise_params}
  \begin{tabular}{c c c c}
    \toprule
    \textbf{Qubit} & \textbf{\(k\) (MHz/\(\Phi_0\))} & \textbf{\(A\) (GHz/\(\Phi_0^2\))} & \textbf{\(\sigma_\Phi\) (\(\mu\Phi_0\))} \\
    \midrule
    Q1 &15.01 & 24.525 & 137.747 \\
    Q2 &3.74 & 27.298 & 30.835 \\
    Q3 &3.98 & 27.089 & 33.049 \\
    Q4 &4.10 & 26.153 & 35.312 \\
    Q5 &2.39 & 26.086 & 20.619 \\
    Q6 &2.33 & 25.783 & 20.348 \\
    Q7 &2.70 & 25.866 & 23.510 \\
    Q8 &3.60 & 24.899 & 32.530 \\
    \bottomrule
  \end{tabular}
\end{table}

Once $k$ is obtained from the fit, the amplitude of the flux noise is calculated by rearranging Eq.~\eqref{eq:k_param}:
\begin{equation}
\sigma_{\Phi} = \frac{\sqrt{2} \ k}{2\pi A}
\label{eq:final_sigma}
\end{equation}
The constant $A$ is determined from independent spectroscopy measurements of the qubit's flux dispersion curve, $f(\Phi) = -\frac{A}{2} \Phi^2 + c$.

This method provides a direct and quantitative measure of the rms flux noise, $\sigma_\Phi$, by leveraging the qubit itself as a highly sensitive spectrometer for its own local magnetic environment. The key signature is the characteristic linear scaling of the dephasing rate with the flux bias detuning, $|\Phi|$, away from the sweet spot. The results presented in Table~\ref{table:flux_noise_params} quantitatively characterize the flux noise environment for each qubit channel, which is critical for understanding dephasing mechanisms and guiding device design and material improvements to achieve longer coherence times, as discussed in Refs.~\cite{Kumar2016, Huang2021}.
 
\section{Supplementary Results}
\subsection{Metrology Performance Across Different Qubit Numbers}

\begin{figure*}
		\begin{minipage}[b]{1.0\textwidth}
			\centering
            \includegraphics[width=0.95\textwidth]{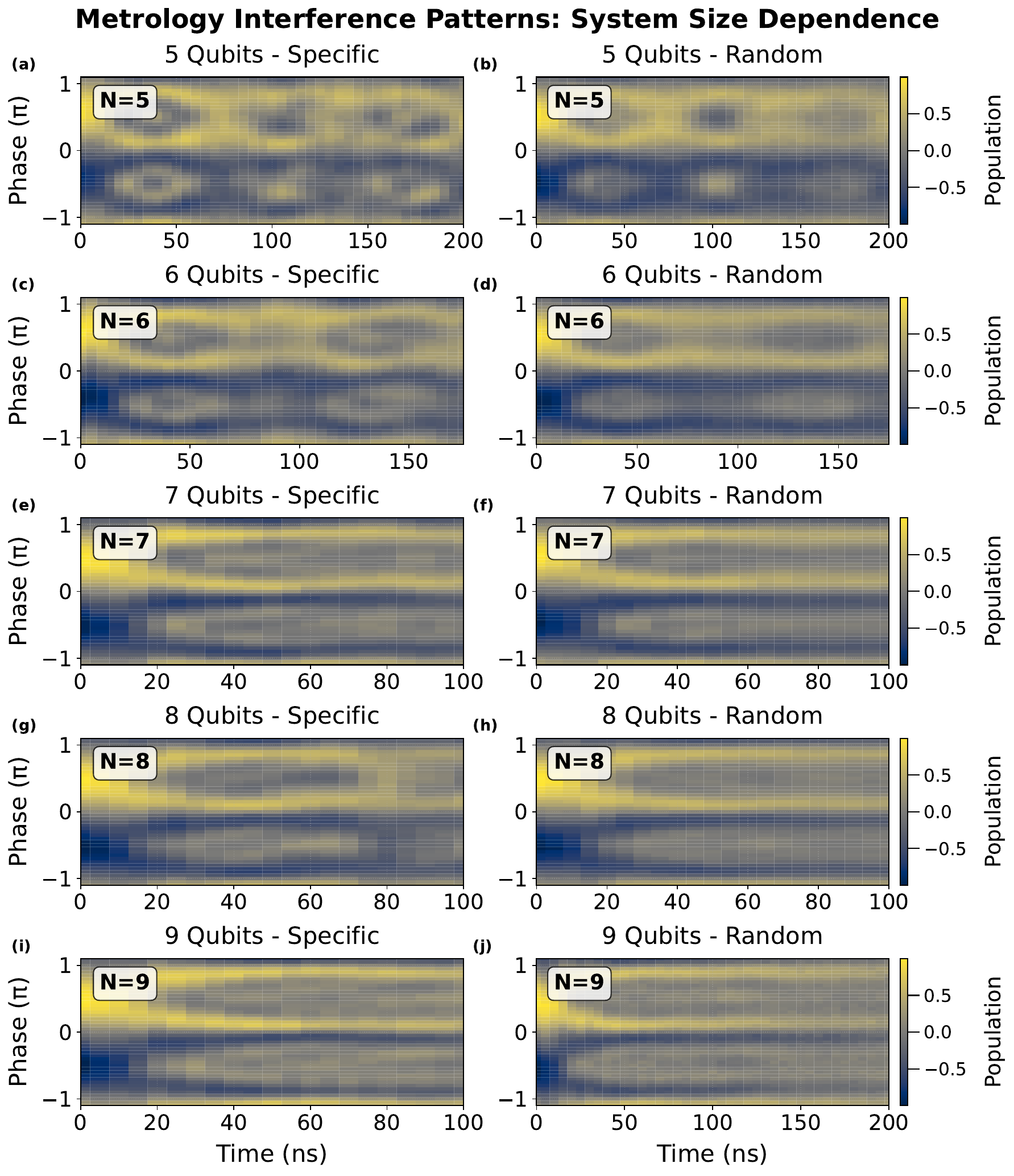}
		\end{minipage}
		\caption{\justifying 
			\textbf{Interference patterns across different qubit numbers.}
			Two-dimensional interference fringes showing the time evolution of quantum phase distributions for \textbf{(a, c, e, g, i)} specific initial states and \textbf{(b, d, f, h, j)} random initial states, for system sizes ranging from 5 to 9 qubits. The horizontal axis represents evolution time (up to $\sim\!200\,\mathrm{ns}$), the vertical axis denotes the encoded quantum phase ($-1.1\pi$ to $1.1\pi$), and color intensity corresponds to the expectation value of $\sigma_x$. Each panel displays the real part of the averaged interference signal, obtained from nine independent experimental runs.
        \label{fig:FIGS5}
			}
\end{figure*}

While the main text presents sensitivity measurements and data analysis for the 9-qubit configuration, we systematically characterize the protocol's performance across varing system sizes. Fig.~\ref{fig:FIGS5} shows the interference patterns ranging from 5 to 9 actively participating qubits. The measured sensitivity scaling demonstrates the expected quantum enhancement with increasing system size.

For each qubit number $N=5,6,7,8,9$, we investigate the protocol under two classes of initial-state configurations: \emph{random} initial states and a \emph{specific} (fixed) initial state. This allows us to assess both the robustness of the protocol against variations in state preparation and the intrinsic metrological performance for a given system size.

In the random initial-state configuration, for each qubit number $N$, we randomly sample nine distinct computational-basis states within the corresponding $N$-qubit active subspace. Each state is represented as a binary occupation vector over the full register, with random amount of qubits in the excited state and the remaining qubits initialized in the ground state. For a given $N$, the sensing experiment is performed independently for each of the nine randomly chosen initial states, and the measured signal and sensitivity are obtained by averaging over these nine realizations. This procedure captures state-dependent fluctuations and provides a representative estimate of the average metrological performance for a given system size.

In the specific initial-state configuration, for each qubit number $N$, we fix a single computational-basis state with $N$ active qubits as a representative initial state. This chosen state is then prepared and measured repeatedly nine times under identical experimental conditions. The reported signal and sensitivity are obtained by averaging over these nine repeated measurements. By keeping the initial-state structure fixed, this configuration isolates statistical fluctuations arising from experimental noise and finite sampling.

The same number of random samples and repeated measurements is used consistently for all qubit numbers from 5 to 9, ensuring a fair comparison across different system sizes. As shown in Fig.~\ref{fig:FIGS5}, both random and specific initial-state configurations exhibit interference patterns and similar sensitivity scaling behavior. This indicates that the observed quantum enhancement is robust with respect to the choice of initial computational-basis states and is primarily governed by the number of actively participating qubits.

\subsection{Gaussian Noise Model for Robustness Testing}
\label{ssec:noise_model}

To characterize the robustness of our information-scrambling enhanced quantum sensing protocol, we introduce Gaussian noise at different stages: time-reversed state preparation and signal sensing. Coherent noise is implemented via Hamiltonian perturbations during the resonant evolution $U(t)$, whereas incoherent noise is modeled as direct perturbations to the probed signal. Specifically, we implement:

\begin{enumerate}

    \item \textbf{Qubit frequency noise:} Gaussian noise $\eta_\omega \sim \mathcal{N}(0,\sigma_\omega^2)$ added to individual qubit frequencies, implemented via flux modulation of each transmon. The perturbed evolution becomes $U_\omega(t) = \exp[-i(H_0 + \sum_{i=1}^9 \eta_\omega^{(i)}(t) a_i^\dagger a_i)t]$.

    \item \textbf{Signal noise:} Incoherent Gaussian noise $\eta_\phi \sim \mathcal{N}(0,\sigma_\phi^2)$ added directly to the probed phase $\phi$, simulating measurement uncertainty and environmental fluctuations during the sensing interval.
\end{enumerate}

The time-reversed state preparation with noise injection takes the form $U^\prime(t)(I + iV)U^\prime(-t)$, where $U^\prime(t)$ represents the noisy forward evolution perturbed by the noise models described above and $U^\prime(-t) = Z \cdot U^\prime(t) \cdot Z$.

For each noise type, we test five strength levels:
\begin{itemize}
    \item Frequency noise: $\sigma_\omega/2\pi = \{0.1, 0.2, 0.3, 0.4, 0.5\}~\mathrm{MHz}$.
    \item Signal noise: $\sigma_\phi = \{0.1, 0.2, 0.3, 0.4, 0.5\}~\mathrm{rad}$.
\end{itemize}

\begin{figure*}
		\begin{minipage}[b]{1.0\textwidth}
			\centering
            \includegraphics[width=0.95\textwidth]{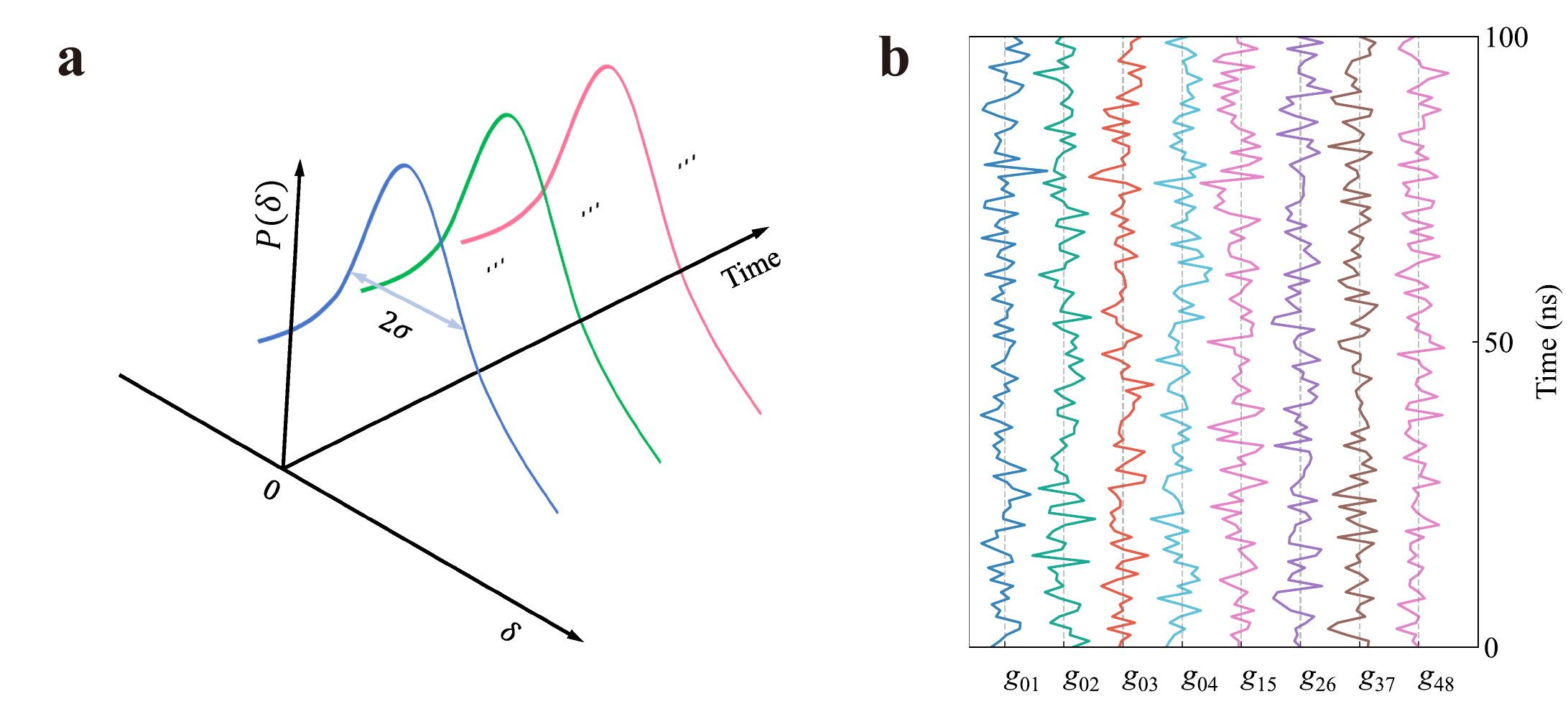}
		\end{minipage}
        \caption{\textbf{Gaussian noise model for robustness testing.} (\textbf{a}) Gaussian probability distributions with standard deviations $\sigma$ corresponding to the five strength levels used for each noise type (coupling, frequency, and signal noise). Each distribution is centred at zero and normalized to unit area. (\textbf{b}) Representative time traces of the coupling noise $\eta_J(t)$ applied to each of the eight inter-qubit couplers over a $100\,\mathrm{ns}$ evolution window. The noise is updated every $\tau_{\mathrm{noise}} = 10\,\mathrm{ns}$, providing a quasi-static perturbation during each segment of the time-reversed state preparation. Coloured lines distinguish individual coupler channels.}
			\label{fig:FIGS6}
\end{figure*}

Fig.~\ref{fig:FIGS6} illustrates our noise implementation. Fig.~\ref{fig:FIGS6}a shows the Gaussian noise distributions for different $\sigma$ values, while Fig.~\ref{fig:FIGS6}b displays representative time traces of the applied coupling noise for the eight inter-qubit couplers during evolution. The noise is updated every $\tau_{\mathrm{noise}} = 10~\mathrm{ns}$, shorter than the system's correlation time but longer than the single-qubit rotation time ($\sim 50~\mathrm{ns}$), ensuring the noise approximates a static disorder during each evolution segment.

Results (presented in Fig.~5 of the main text) demonstrate that our protocol maintains quantum enhancement for noise strengths up to $\sigma_\omega/2\pi \approx 0.3~\mathrm{MHz}$, and $\sigma_\phi \approx 0.2~\mathrm{rad}$, confirming robustness against realistic experimental imperfections.

\bibliography{supp} 